\documentclass[acmtog,authorversion,nonacm]{acmart}
\acmSubmissionID{322}

\usepackage{booktabs} 
\usepackage{mathtools} 
\usepackage{multicol}
\usepackage[ruled]{algorithm2e} 

\SetAlFnt{\small}
\SetAlCapFnt{\small}
\SetAlCapNameFnt{\small}
\SetAlCapHSkip{0pt}

\begin{document}
\title{Differentiable Rendering of Neural SDFs through Reparameterization}
\author{Sai Praveen Bangaru}
\affiliation{%
   \institution{MIT CSAIL}
   \country{USA}
}
\email{sbangaru@mit.edu}

\author{Micha\"el Gharbi}
\affiliation{%
   \institution{Adobe Research}
   \country{USA}
}
\email{mgharbi@adobe.com}

\author{Tzu-Mao Li}
\affiliation{%
   \institution{UC San Diego}
   \country{USA}
}
\email{tzli@ucsd.edu}

\author{Fujun Luan}
\affiliation{%
   \institution{Adobe Research}
   \country{USA}
}
\email{fluan@adobe.com}

\author{Kalyan Sunkavalli}
\affiliation{%
   \institution{Adobe Research}
   \country{USA}
}
\email{sunkaval@adobe.com}

\author{Milo\v{s} Ha\v{s}an}
\affiliation{%
   \institution{Adobe Research}
   \country{USA}
}
\email{mihasan@adobe.com}

\author{Sai Bi}
\affiliation{%
   \institution{Adobe Research}
   \country{USA}
}
\email{sbi@adobe.com}

\author{Zexiang Xu}
\affiliation{%
   \institution{Adobe Research}
   \country{USA}
}
\email{zexu@adobe.com}

\author{Gilbert Bernstein}
\affiliation{%
   \institution{MIT CSAIL \& UC Berkeley}
   \country{USA}
}
\email{gilbo@berkeley.edu}

\author{Fr\'{e}do Durand}
 \affiliation{%
   \institution{MIT CSAIL}
   \country{USA}
}
\email{fredo@mit.edu}

\begin{teaserfigure}
\centering
\includegraphics[width=1\textwidth]{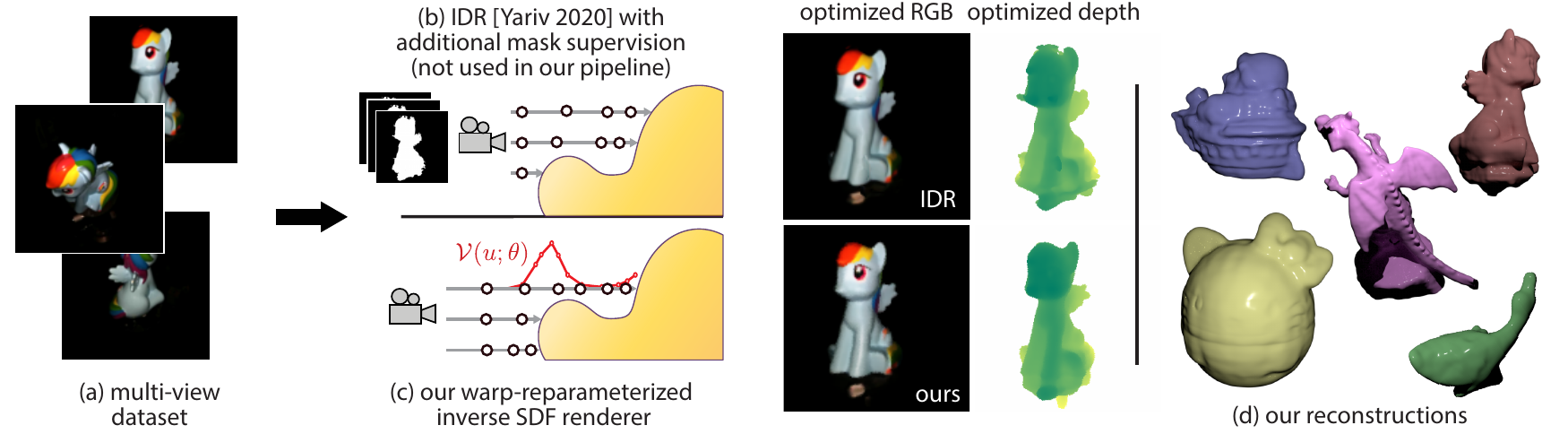}
\vspace{-.25in}
\caption{We propose a novel method to correctly differentiate a neural SDF
renderer by reparameterizing the pixel integral.
%
Direct application of automatic differentiation to the renderer fails because of discontinuities like silhouette boundaries.
In this work we show that, by carefully designing a \textit{discontinuity-aware} warp function $\mathcal{V}(u;\theta)$ to reparameterize the pixel domain, we can remove these discontinuities, and the reparameterized integral is amenable to automatic differentiation.
We demonstrate the benefits of our method on inverse rendering problems.
Starting from a multiview dataset of real photos (a), our reparameterized renderer
(c) can optimize a neural SDF that closely matches the input data, and
generalizes to novel views.
Our renderer matches or outperforms prior SDF renderers~\cite{Yariv:2020:IDR} (b), while doing away with their need for additional geometric supervision in the form of per-view masks, which can be unreliable for real-world data.
We show additional surface reconstructions obtained with our inverse renderer in (d).
%
%
%
}
\label{fig:teaser}
\end{teaserfigure} 

\begin{abstract}
%
%
We present a method to automatically compute \emph{correct} gradients with respect to geometric scene parameters in neural SDF renderers.
Recent physically-based differentiable rendering techniques for meshes have used \emph{edge-sampling} to handle discontinuities, particularly at object silhouettes, but SDFs do not have a simple parametric form amenable to sampling.
Instead, our approach builds on \emph{area-sampling} techniques and develops a continuous warping function for SDFs to account for these discontinuities. 
%
Our method leverages the distance to surface encoded in an SDF and uses quadrature on sphere tracer points to compute this warping function.
We further show that this can be done by subsampling the points to make the method tractable for neural SDFs.
Our differentiable renderer can be used to optimize neural shapes from multi-view images and produces comparable 3D reconstructions to recent SDF-based inverse rendering methods, without the need for 2D segmentation masks to guide the geometry optimization and no volumetric approximations to the geometry.
\end{abstract}

%
%

%
%


\maketitle

\section{Introduction}

%
Differentiable rendering algorithms have become crucial tools in solving
challenging inverse problems~\cite{Zhao:2020:PSDR}, thanks to their ability to
compute the derivatives of images with respect to arbitrary scene
parameters.
Naive differentiation of rendering algorithms does not handle discontinuities caused by visibility changes and object boundaries correctly.
Previous work has observed that the discontinuities can be handled by properly handling the Dirac delta signals, and derived algorithms for explicit geometry representations like triangle meshes~\cite{Li:2018:DMC,Zhang:2020:PSDR}.
%
%

%
On the other hand, implicit representations like signed distance fields (SDFs) are appealing since they do not require the initialized geometry to have the right topology. 
%
%
Recent work has demonstrated the use of SDFs---usually parameterized using multi-layer perceptron networks---for the task of reconstructing shape and surface reflectance from images.
However, these methods either require additional geometric supervision such as segmentation masks~\cite{Yariv:2020:IDR,Zhang:2021:PIR} or make approximations to the geometry using volumetric models~\cite{yariv2021volume,Oechsle2021ICCV} that limit their applicability.
%
%
%
%
%
%

%
%
In this paper, we derive an algorithm to automatically compute \emph{correct} gradients with respect to geometric scene parameters in neural SDF renderers.
Previous methods that rely on silhouette sampling are not directly applicable to SDFs since direct sampling of boundaries of implicit functions is challenging.
Instead, we build on the \emph{reparameterization} approaches~\cite{Loubet:2019:RDI,Bangaru:2020:WAS}, which removes discontinuities through reparameterization while preserving the integral values. 
These methods do not require explicit sampling along discontinuities.
Previous reparameterization methods focused on triangle meshes, and require new derivation for reparameterizing SDF rendering. 
%

\begin{figure*}[!t]
\centering
\includegraphics[width=\textwidth]{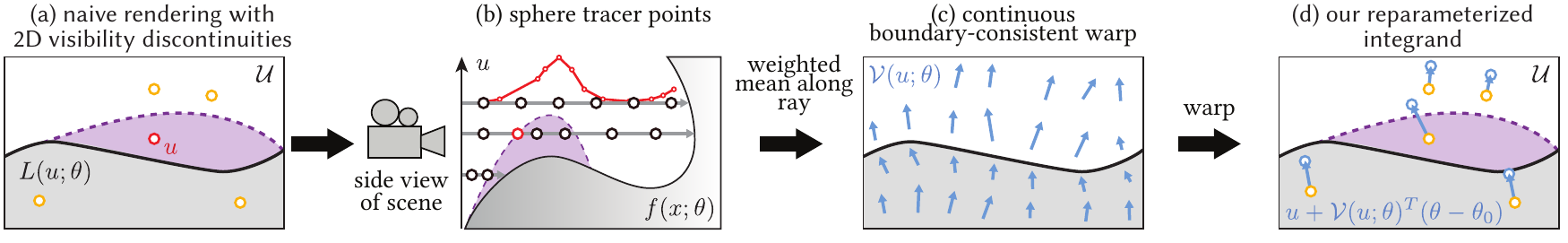}
\vspace{-.25in}
\caption{\label{fig:overview}\textsc{Overview}.
Without proper case a SDF rendering pipeline is discontinuous, which means there are points $u$ where the rendering function $L(u;\theta)$ is not differentiable in $\theta$, highlighted in red (a). 
Our method uses intermediate points from a sphere tracer (b) applied to an SDF $f$, to compute a warp function $\mathcal{V}$ (c).
Using this warp, we reparameterize the integration domain to avoid discontinuities (d), which
allows us to compute correct gradients of the rendering equation. 
The key to achieving this is to design the warp $\mathcal{V}$ so it is continuous in $u$ everywhere, and satisfies some consistency criterion on the geometric boundaries.
}
\end{figure*}

%
%
Specifically, we construct a silhouette-aware reparameterization similar to that of \citet{Loubet:2019:RDI}, but following the equivalent unbiased warp definition that \citet{Bangaru:2020:WAS} used to produce correct gradients for triangle-meshes.
We leverage the fact that SDFs naturally encode the distance to the surface, and develop a practical algorithm that uses a quadrature on sphere tracing~\cite{Hart:1996:STG} samples to construct a reparameterization that removes the discontinuities.
We further show that this can be computed using only a subset of sphere tracing samples, reducing computational burden of the backward pass for bulky neural SDFs\footnote{We only became aware of the concurrent work "Differentiable Signed Distance Function Rendering" [Vicini et al. 2022] a week before submission.
They use a warping reparameterization similar to ours that also builds on~\citet{Bangaru:2020:WAS}, but focus on voxelized SDFs. We show that reparameterization works with neural SDFs.
Our method also uses a more efficient `top-k' weighting scheme.}.


Our algorithm produces correct geometry gradients for SDFs.
It does away with the segmentation masks and depth guidance required by previous techniques~\cite{Yariv:2020:IDR}, without making a volumetric approximation to the geometry~\cite{yariv2021volume,Oechsle2021ICCV}.
We show that our differentiable renderer can be used to optimize neural shapes from multi-view images, with no additional information beyond the RGB data and the corresponding camera parameters.
%
%
Our focus is on occlusion discontinuities, so the rest of the paper assumes a differentiable shading model.


\section{Related Work}
We focus on work that recover the latent 3D scene from images through differentiable rendering. We categorize them by the type of scene representation.

\paragraph{Meshes.} To account for discontinuities, earlier work focused on approximating the derivatives of mesh rendering by smoothing the geometry~\cite{De:2011:M3H,Loper:2014:OAD,Rhodin:2015:VSM,Kato:2018:N3M,Liu:2019:SRD}.
Alternatively, some work derived correct analytical derivatives under simplified assumptions~\cite{Arvo:1994:IJP,Zhou:2021:VFA}.
Li et al.~\shortcite{Li:2018:DMC} noticed that the differentiation of discontinuities caused by the visibility and geometric boundaries lead to Dirac delta signals, and can be integrated by the pixel antialiasing integral or the rendering equation.
They proposed an \emph{edge sampling} algorithm to explicitly sample the Dirac delta on triangle mesh silhouettes.
Importance sampling the silhouettes can be difficult, therefore Loubet et al. and Bangaru et al.~\shortcite{Loubet:2019:RDI,Bangaru:2020:WAS} later proposed to convert the silhouette integral into an area integral.
Loubet et al. formulated the conversion using a reparametrization, and derived an approximated reparametrization to remove discontinuities.
Bangaru et al. built on Loubet et al.'s work and derived an unbiased estimator by showing the equivalence between the reparametrization and divergence theorem.
On the other hand, Zhang et al.~\shortcite{Zhang:2020:PSDR} showed that directly sampling silhouette in path-space~\cite{Veach:1998:RMC} can also be done efficiently. 
Directly sampling the silhouette for SDFs is difficult.
Our work extends the reparametrization approach to handle SDFs, including approximate SDFs defined by neural networks.

\paragraph{Level sets and signed distance fields.} A level set defines a surface using the roots of a 3D implicit function. A signed distance field is a specific kind of level set where the implicit function defines the distance of a 3D point to the surfaces, where the sign is negative when the point is inside the object. SDFs can be represented using polynomials~\cite{Blinn:1982:GAS}, voxels~\cite{Izadi:2011:KFR}, or neural networks~\cite{Park:2019:DLC}. 
Differentiable rendering for SDFs has been discussed in computer vision and used for 3D surface reconstruction~\cite{Niemeyer:2020:DVR,Jiang:2020:SDR,Yariv:2020:IDR,Zhang:2021:PIR,Kellnhofer:2021:NLR}, but currently methods all ignore the discontinuities when differentiating and require 2D object masks to converge.
An alternative way to render the signed distance field is to convert it to another format such as a thin participating medium~\cite{Oechsle2021ICCV,yariv2021volume,Wang:2021:NLN}, a mesh~\cite{Remelli:2020:MDI}, or a point cloud~\cite{Cole:2021:DSR}. These methods all introduce approximation. Instead, we focus on deriving accurate gradients without approximation.

\paragraph{Volumes.} A scene can also be represented as participating media instead of solid surfaces. Gkioulekas et al.~\shortcite{Gkioulekas:2013:IVR} pioneered the use of differentiable volume rendering for inverse problems. Zhang et al.~\shortcite{Zhang:2019:DTR,Zhang:2021:PSD} tackled discontinuities at volumetric boundaries. Recently, there has been a surging interest in using volumetric representations---parameterized either as discretized grids or neural networks---for view synthesis~\cite{Lombardi:2019:NV,Mildenhall:2020:NRS,Liu:2020:NSVF,Xie:2022:NFV}. These volumetric representations allow for a trivially differentiable rendering model and can achieve high-quality novel view synthesis and appearance acquisition~\cite{Bi:2020:DRV,Bi:2020:NRF}. However, it is still a challenge to extract high-quality surface geometry from these methods, and while the trade-offs between surface and volume representations is an interesting research topic, we focus on surface representations.

\paragraph{Light transport.} In addition to handling discontinuities, recent work also studies the reduction of variance and memory consumption for Monte Carlo rendering~\cite{Zeltner:2021:MCE,Zhang:2021:ASM,NimierDavid:2020:Radiative,Vicini:2021:PRB}. Earlier rendering work used derivatives for forward rendering~\cite{Ward:1992:IG,Ramamoorthi:2007:FAL,Li:2015:AGM,Luan:2020:LMC}. Our work is largely orthogonal to these.

\section{Method}
Our method computes the correct gradient of a rendering function (i.e. the pixel integral of the radiance function on the camera image plane) with respect to geometric parameters, in the presence of primary visibility discontinuities, for scenes where the geometry
is represented by a signed distance field $f$, parameterized by $\theta$ (e.g., the weights of neural network).
Our approach builds on~\citet{Bangaru:2020:WAS}.
We show how to extend their warp function to SDFs in order to reparameterize an intractable boundary integral.
We summarize the necessary background in \S~\ref{sec:background}.
We then derive a warp function for SDFs that is continuous and boundary consistent (\S~\ref{sec:sdf_reparam}) as an integral along camera rays, and show how to compute it via quadrature using sphere tracer points (\S~\ref{sec:quadrature}).
In Section~\ref{sec:topk_approx}, we finally give an unbiased approximation for this warp that is tractable for use with neural SDFs, and we show the criteria of unbiasedness and a sketch of proof.
Section~\ref{sec:inverse_rendering} provides details on how to use our approach to solve inverse rendering problems.

\subsection{Background: boundary-aware warping}\label{sec:background}

Without loss of generality, assume a box pixel filter, so that $\mathcal{U}\subset\mathbb{R}^2$ is the image plane region corresponding to the pixel of interest. Let $L(u; \theta)$ denote the radiance along the ray from $u\in\mathcal{U}$, a point on the image plane, and denote $\theta\in\mathbb{R}^N$ the vector of geometric scene parameters (e.g. neural network weights). In matrix expressions below, we will assume vector quantities ($u$, $x$, $\theta$) to be row vectors, and gradients with respect to $\theta$ to be column vectors.

\begin{figure}[!tb]
    \centering
    \includegraphics[width=\columnwidth]{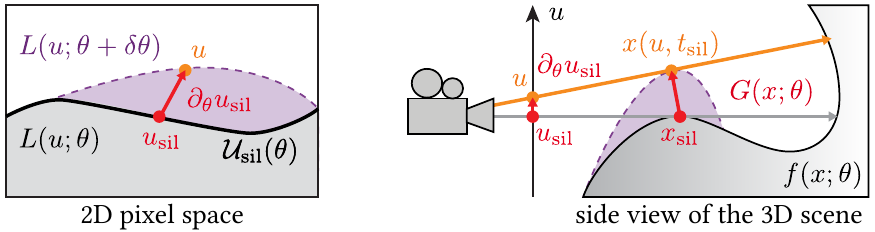}
    \vspace{-.25in}
    \caption{\label{fig:du-dtheta}
    As geometric parameters $\theta$ vary, visibility creates discontinuities in the rendering function $L(u;\theta)$ which traditional automatic differentiation cannot handle (left).
    These discontinuities correspond to continuous changes in our SDF representation $f(x;\theta)$ (right).
    We compute the derivative $\partial_\theta u_{sil}$ of a pixel-space silhouette point w.r.t.\ the geometry parameters $\theta$ by computing the derivative $G(x;\theta)$ of the corresponding 3D scene point $x$, and projecting it onto the screen space $\mathcal{U}$ through the inverse Jacobian.}
\end{figure}

We aim to compute the gradient of the rendering integral $I$ with respect to parameters $\theta$:
\begin{equation}
   \partial_\theta I = 
   \frac{\partial}{\partial\theta}\int_\mathcal{U} L(u; \theta)du.
   \label{eqn:rendering_derivative}
\end{equation}
Primary visibility discontinuities make the radiance function non-differentiable along occlusion boundaries (Fig.~\ref{fig:du-dtheta}).
Denoting $\mathcal{U}_{\text{sil}}(\theta) \subset \mathcal{U}$ the set of
object silhouettes, for a point $u_{\text{sil}} \in\mathcal{U}_{\text{sil}}$,
the radiance $L(u_{\text{sil}}; \theta)$ is discontinuous in
$\theta$.
This makes naive automatic differentiation methods applied to the Monte Carlo sampling of $I$ produce incorrect gradients since they ignore the Dirac delta that arises from the differentiation.

Li et al.\shortcite{Li:2018:DMC,Li:2019:DVC} and \citet{Zhang:2019:DTR} showed that Eq.~\eqref{eqn:rendering_derivative} can 
be split into two terms: an interior integral, for contributions away from the discontinuities; and a boundary integral, along the discontinuities:
\begin{equation}
   \partial_\theta I = 
   \int_\mathcal{U} \frac{\partial}{\partial\theta}L(u; \theta)du + I_{\text{sil}}.
   \label{eqn:rendering_derivative_partition}
\end{equation}
%
The second integral $I_{\text{sil}}$ is harder to compute because sampling the boundary is generally technically difficult. This is particularly true for SDFs whose surface boundaries admit no easy parametric form. We will not cover boundary sampling in detail, since we will not use it; instead, we will use a result from 
\citet{Bangaru:2020:WAS}, who showed, using the divergence theorem, that this boundary term can be turned into an integral over the interior $\mathcal{U}\setminus\mathcal{U}_{\text{sil}}(\theta)$, which is easier to sample:
\begin{equation}
   I_{\text{sil}} = 
   \int_{\mathcal{U}\setminus\mathcal{U}_{\text{sil}}(\theta)} \nabla_u \cdot \left(L(u; \theta)\mathcal{V}(u;\theta)\right)du.
   \label{eqn:divergence_theorem}
\end{equation}
Here $\nabla_u \cdot$ is the divergence operator,
and $\mathcal{V}(u;\theta)\in\mathbb{R}^{N\times 2}$ is a warping function required to satisfy two properties:
\begin{enumerate}
    \item \textbf{continuity}: $\mathcal{V}(\cdot;\theta)$ is continuous on $\mathcal{U}$, and
    \item \textbf{boundary consistency}: $\mathcal{V}$ agrees with the derivative of the discontinuity points when $u$ approaches the discontinuity.
    That is, $\lim_{u \to u_{\text{sil}}}\mathcal{V}(u; \theta) = \partial_{\theta}u_{\text{sil}}$ for $u_\text{sil}\in \mathcal{U}_\text{sil}(\theta)$.
\end{enumerate}
Bangaru et al. further show that the area integral is equivalent to applying the change of variable~\cite{Loubet:2019:RDI} 
$u \mapsto T(u,\theta) = u + (\theta - \theta_0) \mathcal{V}(u;\theta)$
 in Eq.~\eqref{eqn:rendering_derivative}, where $\theta_0$ is the point at which the derivative is computed but $\partial_\theta \theta_0 = 0$.
Applying the reparameterization we obtain:
\begin{align}
  \partial_\theta I = \int_{\mathcal{U}} \partial_\theta \left[L(T(u,\theta),\theta) \left|\det\left(\partial_u T(u,\theta)\right)\right|\right] \mathrm{d}u.
  \label{eqn:final_integral}
\end{align}
Expanding $T$ and using Eq.~\eqref{eqn:divergence_theorem}, one can show that Eq.~\eqref{eqn:final_integral}  indeed computes $\partial_\theta I$. 
Intuitively, the reparameterization $T$ moves each point on the boundary locally at the velocity of their derivatives, essentially removing the discontinuities, while the determinant term accounts for the change of measure. 

The main goal of this paper is to derive a suitable form for $\mathcal{V}(u;\theta)$ for SDFs, that can be tractably computed, so that we can evaluate Eq.~\eqref{eqn:final_integral} using Monte Carlo estimation.
%
%

\paragraph{Rendering.}
To render an SDF $f$ and compute $L(u;\theta)$, we need to find the closest intersection point $x(u,t) \in \mathbb{R}^3$ such that $f(x;\theta) = 0$, where $t$ is the distance along the primary ray associated with pixel location $u$. To find the intersection distance, we use sphere tracing~\cite{Hart:1996:STG}, which applies a fixed-point iteration to generate a sequence of points $x_n \in \mathcal{T}(u)$, such that $\lim_{n\rightarrow\infty} x_n = x$.

\subsection{Continuous boundary-consistent warp for SDFs}\label{sec:sdf_reparam}
In this section, we construct an idealized warp function $\mathcal{V}^{\text{int}}$ that satisfies the continuity and boundary-consistency conditions of Section~\ref{sec:background}.
%
%
First, we derive the boundary gradient $\partial_\theta u_\text{sil}$ with which the warp should agree at silhouette points (\S~\ref{sec:boundary_consistency}).
We then smoothly extrapolate this gradient using a weighted integral along the primary ray passing through $u$, to obtain our warp function (\S~\ref{sec:ideal_warp}).
We show necessary and sufficient conditions on the weighs to make the warp continuous and boundary-consistent (\S~\ref{sec:ideal_weights}).

\subsubsection{Boundary consistency for implicit functions}\label{sec:boundary_consistency}

The boundary consistency condition in Section~\ref{sec:background}, requires that, at a discontinuity point $u_{\text{sil}}$ the warp agrees with $\partial_{\theta} u_\text{sil}$.
The derivation proposed by \citet{Bangaru:2020:WAS} does not apply directly to implicit surfaces, so
we derive this boundary derivative using the implicit function theorem.
Specifically, the derivative of a scene point $x\in\mathbb{R}^3$ on the surface, i.e., $f(x;\theta) = 0$, w.r.t.\ parameters $\theta\in\mathbb{R}^N$ is given by:
\begin{equation}
   G(x; \theta) \vcentcolon= \partial_{\theta} x = -\frac{\partial_{\theta} f \partial_{x} f^T}{\|\partial_{x} f\|^2} \in \mathbb{R}^{N\times 3}.
\end{equation}
The above directly follows from the implicit function theorem applied to $f(x;\theta) = 0$.
To get the derivative in pixel coordinates $\partial_\theta u = \partial_\theta x \cdot \partial_x u\in\mathbb{R}^{N\times 2}$, we need to project this derivative by the Jacobian $\partial_{x}u\in\mathbb{R}^{3\times 2}$, 
which for a perspective camera can be easily derived by hand. For more generality, 
%
%
we can obtain this Jacobian as the pseudo-inverse $\dagger$ of the forward Jacobian:
\begin{equation}
    \partial_{x}u = (\partial_u x(u, t))^{\dagger}.
\end{equation}
%

Taken together, the derivative at a silhouette point $u_{\text{sil}}$, with corresponding 3D position $x_\text{sil} = x(u_{\text{sil}}, t_{\text{sil}})$, is then:
\begin{equation}
    \partial_{\theta} u_\text{sil} = G(x_{\text{sil}}; \theta) \, \partial_{x} u.
    \label{eq:boundary_consistency}
\end{equation}
%
Figure~\ref{fig:du-dtheta} illustrates the geometric configuration.

\subsubsection{Extending to a smooth warp $\mathcal{V}^{\text{int}}(u;\theta)$ by integration along the ray}\label{sec:ideal_warp}

\begin{figure}[!tb]
    \centering
    \includegraphics[width=\columnwidth]{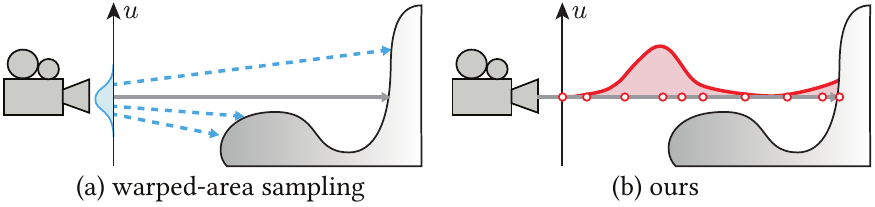}
    \vspace{-.25in}
    \caption{\label{fig:was-illustration}(a) Warped-area sampling uses additional random rays around the primal ray and averages $\partial_{\theta}{x(u;t)}$ using boundary-aware harmonic weights. (b) Our method instead takes a weighted average along the ray, repurposing the SDF $f$ into weights $S(x)^{-\gamma}$.}
\end{figure}

Now that we have an expression for the warp at silhouette points, we extend it to all points, by smoothing this term in a consistent manner. 
%
%
Our method takes advantage of the fact that our implicit SDF $f(x;\theta)$ is continuous in 3D space and achieves the smoothing by convolving \emph{along the ray} (Fig.~\ref{fig:was-illustration} (b)).
This avoids casting expensive additional rays which are needed by \citet{Bangaru:2020:WAS}, and also propagates gradients to points in free space \textit{near} the boundary points.
This can have a stabilizing effect on the optimization of neural SDFs, as noted by \citet{Oechsle2021ICCV} and \citet{Wang:2021:NLN}; note that, while they adapt a volumetric rendering model to achieve better convergence, we do so while computing correct boundary gradients for a surface-based representation.
%

Our proposed warp function smoothly extends Eq.~\eqref{eq:boundary_consistency} to non-boundary points as follows:
\begin{equation}
    \mathcal{V}^{\text{int}}(u;\theta) = 
    \frac{\int^{t_0}_{t=0} w\left(x(u,t)\right)\,
     G(x;\theta) \, \partial_{x}u \, dt}
    { \int^{t_0}_{t=0} w\left(x(u,t)\right) \, dt},
    \label{eqn:v-integral}
\end{equation}
with $t_0$ the distance to the closest intersection, $t_0=\infty$ when the ray does not intersect.

\subsubsection{Choice of weights}\label{sec:ideal_weights}

In order to satisfy the boundary consistency criteria, the weights need to asymptotically satisfy the limit:
\begin{equation}
    \lim_{u \to u_{\text{sil}}} \frac{w\left(x(u,t)\right)}{\int^{t_0}_{t^{\prime}=0} w\left(x(u,t^{\prime})\right) \, dt^{\prime}} = 
    \delta(t - t_{\text{sil}}),
    \label{eqn:wt-limit}
\end{equation}
where $\delta$  is the Dirac delta operator.

From Eq.~\eqref{eqn:wt-limit}, we see that our weights have to depend on some notion of distance to the silhouette.
For an implicit function $f$ that is at least $C_1$ continuous, the following constraints implicitly characterize the silhouette points~\cite{Gargallo:2007:MRE}:  
\begin{equation}
\begin{aligned}
  &f(x(u,t);\theta) = 0, \\
  &\partial_{x} f(x(u,t); \theta)^T\partial_{t} x(u, t) = 0.
\end{aligned}
\end{equation}
The first condition requires the point to be on the surface, and the second condition requires the SDF gradient to be perpendicular to ray direction~\cite{Hertzmann:1999:INP}.
We can use these equations to build a silhouette characteristic function $\mathcal{S}(x)$, which takes value $0$ whenever $x$ is a silhouette point, and is continuous everywhere.
Specifically, we define:
\begin{equation}
    \mathcal{S}(x) = \left|f(x; \theta)\right| + \lambda_d \, \left|\partial_{x} f(x; \theta)^T\partial_{t} x\right|,
\end{equation}
where $\lambda_d > 0$.
This characteristic function is similar to the boundary test function used by \citet{Bangaru:2020:WAS} for meshes.
However, unlike their boundary test, $\mathcal{S}(x)$ is defined everywhere in the SDF's 3D domain, not just the surface points.
This allows us to use these weights for our integral along any ray.

Our final harmonic weights are given by:
\begin{equation}
    w(x) = \mathcal{S}(x)^{-\gamma}, \gamma > 2.
    \label{eqn:harmonic-wts}
\end{equation}
For $\gamma > 2$, our weights satisfy the limit in Eq.~\eqref{eqn:wt-limit}.
Intuitively, this is because the $w(x) \to \delta(x - x_{\text{sil}})$ as $x \to x_{\text{sil}}$.
%
See our supplementary material for a discussion of correctness, and derivation of $\gamma > 2$.

\begin{figure*}[!thb]
    \centering
    \includegraphics[width=\textwidth]{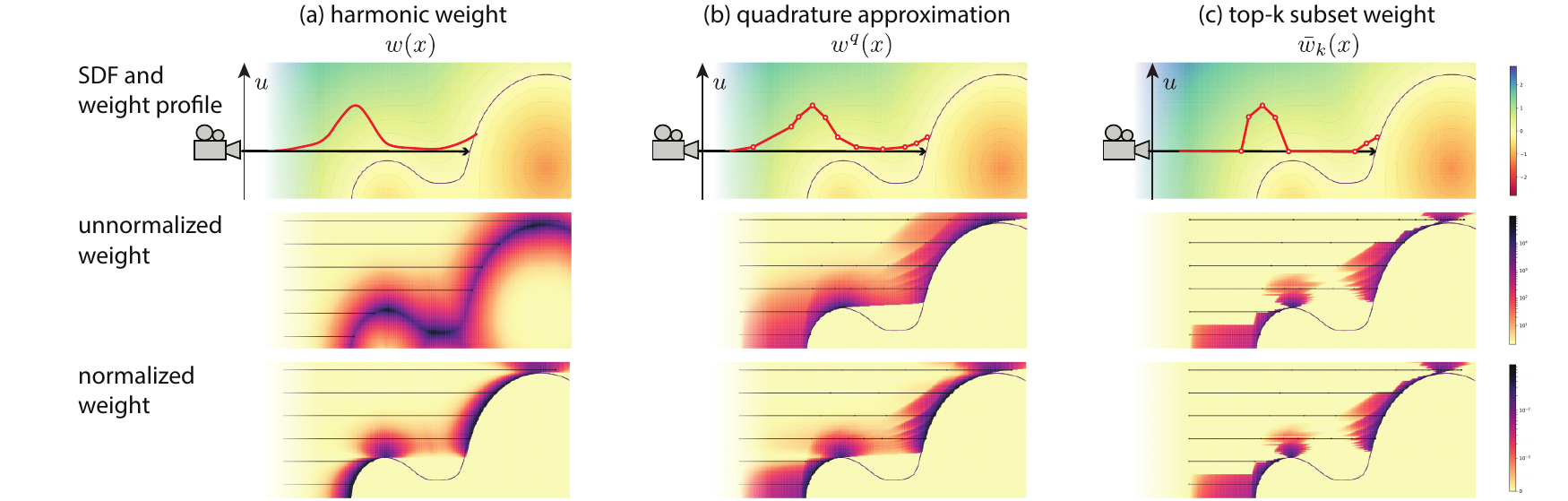}
    \caption{\label{fig:weights}\textbf{Weight visualization.}
    A contour plot of a sample 2D SDF (first row). We use an orthographic camera for illustration, so camera rays are parallel to the horizontal axis.
    We show our three weighting schemes in unnormalized (second row) and normalized (third row) form.
    Our proposed harmonic weights (a) for $\gamma=4.0$, $\lambda_d=1e-1$ are
    well approximated by a trapezoidal quadrature on the sphere tracer points (b).
    The blank regions with no weight can be excluded from the computation,
    which leads to our proposed our proposed top-$k$ subset weights (c),  for $k=8$.
    This reduces both the compute and memory burden of the backward pass.
    %
    %
    %
    We visualize the weight in a \emph{symlog} plot, values are linear in $[0, 10^1]$ and $[0, 10^{-3}]$ for the unnormalized and normalized weights, respectively.}
\end{figure*}

Fig. \ref{fig:weights}(a) shows our weight distribution along the ray for all $u$ in a 1D example sphere tracer.

\subsection{Estimating the warp through its quadrature $\mathcal{V}^{q}$}\label{sec:quadrature}
We now have a clear form for our warp function that can be used to reparameterize and differentiate the rendering function.
Unfortunately, the asymptotical sharpness of our weights required to obtain a valid warp, also makes the integral~\eqref{eqn:v-integral} very difficult to sample.
For $u$ close to the silhouette $\mathcal{U}_{\text{sil}}$, the weights become very concentrated near the surface boundary, presenting a tricky integrand if we were to uniformly sample along the ray.

Careful importance sampling of areas near the boundary could remedy this, but there is unfortunately no straightforward way to implement this: the weight distribution depends heavily on the configuration of silhouettes near $u$, dictated by the SDF.

Our approach foregoes stochastic sampling altogether.
We construct a trapezoidal \textit{quadrature} on the series of intermediate points $x_n\in\mathcal{T}(u)$ generated by the sphere tracer, shown in Fig.~\ref{fig:weights}(b).
This quadrature estimator for the warp is given by:
\begin{equation}
\begin{aligned}
    &\mathcal{V}^{q}(u; \theta) =
    \frac{
        \sum_{x_i \in \mathcal{T}(u)} w^{q}\left(x_i\right) \,
         G(x; \theta) \, \partial_{x}u}
    { \sum_{x_i \in \mathcal{T}(u)} w^{q}\left(x_i\right)}, \\
    &\text{where } w^{q}(x_i) = w(x_i) \frac{(t_{i - 1} - t_{i + 1})}{2},
\end{aligned}
\label{eqn:warp-q}
\end{equation}
and $t_i$ is the distance along the ray to sphere tracer point $x_i$.
Assuming the underlying SDF $f(x;\theta)$ is $C_1$ continuous, the intermediate points of the sphere tracer are continuous at all $u \not\in \mathcal{U}_{\text{sil}}$.
%
%
By composition of continuous functions, $\mathcal{V}^{q}(\cdot; \theta)$ is also continuous.

Our quadrature warp $\mathcal{V}^q$ satisfies the continuity and boundary consistency condition (\S\ref{sec:background}).
Since we apply trapezoidal quadrature, $\mathcal{V}^{q}(u;\theta)$ is in general a biased estimator of integral $\mathcal{V}^{\text{int}}(u;\theta)$.
%
However, the two terms \textit{are} equal in the limit as $u$ approaches the silhouette, i.e., for $u_\text{sil}\in\mathcal{U}_\text{sil}$,
%
    $\lim_{u \to u_{\text{sil}}} \mathcal{V}^{q}(u; \theta) = \lim_{u \to u_{\text{sil}}} \mathcal{V}^{\text{int}}(u; \theta)$,
%
and since the right-hand side is boundary consistent, so is our quadrature warp $\mathcal{V}^{q}$.
See supplemental for a sketch proof of correctness.

\subsection{Top-$k$ subset weighting $\bar{w}_k$ to reduce memory use}\label{sec:topk_approx}
For complex SDFs such as a neural network,
our quadrature warp $\mathcal{V}^{\text{q}}$ has the caveat that it requires back-propagating through \emph{every} sphere tracer point.
Previous work like IDR \cite{Yariv:2020:IDR} do not have this issue since their (biased) gradient is only computed at the intersection point, and they exclude other points from the gradient computation. 
Our approach, on the other hand, uses a weighted sum, so we cannot discard intermediate points.

%

However, as shown in Fig.~\ref{fig:weights}(b), the vast majority of sphere tracer points have negligible weight, and most of the mass is concentrated close to the \textit{silhouette}. 
We exploit this by only using the subset of points with the highest weight in our warp estimation.
That is, instead of using all of $T(u)$, we can instead use a top-k subset $T_k(u)$.
%
%
Selecting the top-$k$ weights requires adjusting them to ensure that they remain continuous.
For a subset size of $k$, our weights are\footnote{Note that even though we consider the top $k$ weights, only $k - 1$ weights actually have a non-zero contribution to the $\theta$-derivative.}:
\begin{equation}
 \bar{w}_k(x) = 
 \begin{cases}
     w^{\text{q}}(x) - \min\limits_{x \in T_k(u)}w^{\text{q}}(x), & \text{if } x \in T_k(u)\\
    0              & \text{otherwise.}
\end{cases}
\end{equation}

The weights $\bar{w}_k(x)$ still produce a continuous warp field (see supplemental for a sketch of proof).
Intuitively, even though the set of points change as a function of $u$, whenever this change occurs, the points that swap in or out of the set always have weight 0. 

\subsection{Inverse Rendering Details}\label{sec:inverse_rendering}
In this section, we briefly discuss some details that make our inverse rendering pipeline tractable.

\paragraph{Implementation}
Our method requires 3 \textit{nested} derivative passes to (i) compute normals $\partial_{\mathbf{x}}f$, (ii) compute Jacobian of the transformation $\partial_u T$ and (iii) to compute derivatives of the full pipeline $\partial_{\theta} \left[L(T(u, \theta)) \left|\det\left(\partial_u T(u, \theta)\right)\right|\right]$.
We use the Python \textbf{JAX} automatic differentiation system~\cite{jax2018github}, which supports nested forward+backward differentiation. We use forward-mode for (i) and (ii), and reverse-mode for (iii). 

\paragraph{Network architecture}
For our inverse rendering results, we use the network architecture shown in Fig.~11 of \citet{Yariv:2020:IDR}. 
Since our method is slightly more memory-intensive (even with top-$k$ subset weights), we reduce the width of the SDF network to 256 channels per layer.
In this architecture, the shading network predicts the final radiance based on the position, viewing direction and a geometric feature vector.
However, in contrast to NeRF~\cite{Mildenhall:2020:NRS}-like methods, the shading network is only evaluated at \emph{surface} points.
We use 6-levels of positional encoding on the input position $x$ to allow the network to reconstruct fine geometry.

\paragraph{Pixel sampling}
Similar to \citet{Yariv:2020:IDR} and other neural methods, we sample a subset of pixels for each iteration since it can be computationally prohibitive to trace the entire image when using a deep neural representation.
However, unlike \citet{Yariv:2020:IDR}, which works with a single ray at the center of the pixel, our approach must integrate the spatially-varying warp $\mathcal{V}$ over each pixel.
We achieve this by Monte-Carlo sampling within each pixel. 
Appendix~\ref{sec:pixel_boundary_sampling} discusses how we incorporate pixel filters.

\paragraph{Multi-level optimization}
Since we only use a subset of pixels, the likelihood of sampling a pixel with silhouette gradient is fairly low.
For unbiased derivatives, only pixels that are partially covered by a surface have a non-zero boundary contribution.
This is in contrast to approximate derivatives (e.g.,~\cite{Liu:2019:SRD},~\cite{Yariv:2020:IDR}) that have a wider spatial footprint.
To alleviate this issue, we use a multi-scale pyramid of the target image throughout our optimization to help with gradient stability.

\paragraph{Initialization}
We use the geometric network initialization~\cite{Atzmon:2020:CVPR} which approximately produces a spherical SDF.
We also initializes the weights of the positional encoding layer to 0~\cite{Yariv:2020:IDR}.
We found this subtle modification implicitly enforces a coarse-to-fine mechanism that yields significantly better generalization to novel views.

\paragraph{Eikonal constraint}
We represent our SDF $f$ using a neural network, which does not necessarily satisfy the distance property. We adopt the Eikonal regularization loss~\cite{Gropp:2020:IGR} to explicitly enforce this.
In-spite of the additional loss, $f$ is still an approximation of an SDF, and therefore we pad our weights with a small $\epsilon$ in order to avoid infinities.
\section{Results}

\begin{figure}[!tbh]
\centering
\includegraphics[width=\columnwidth]{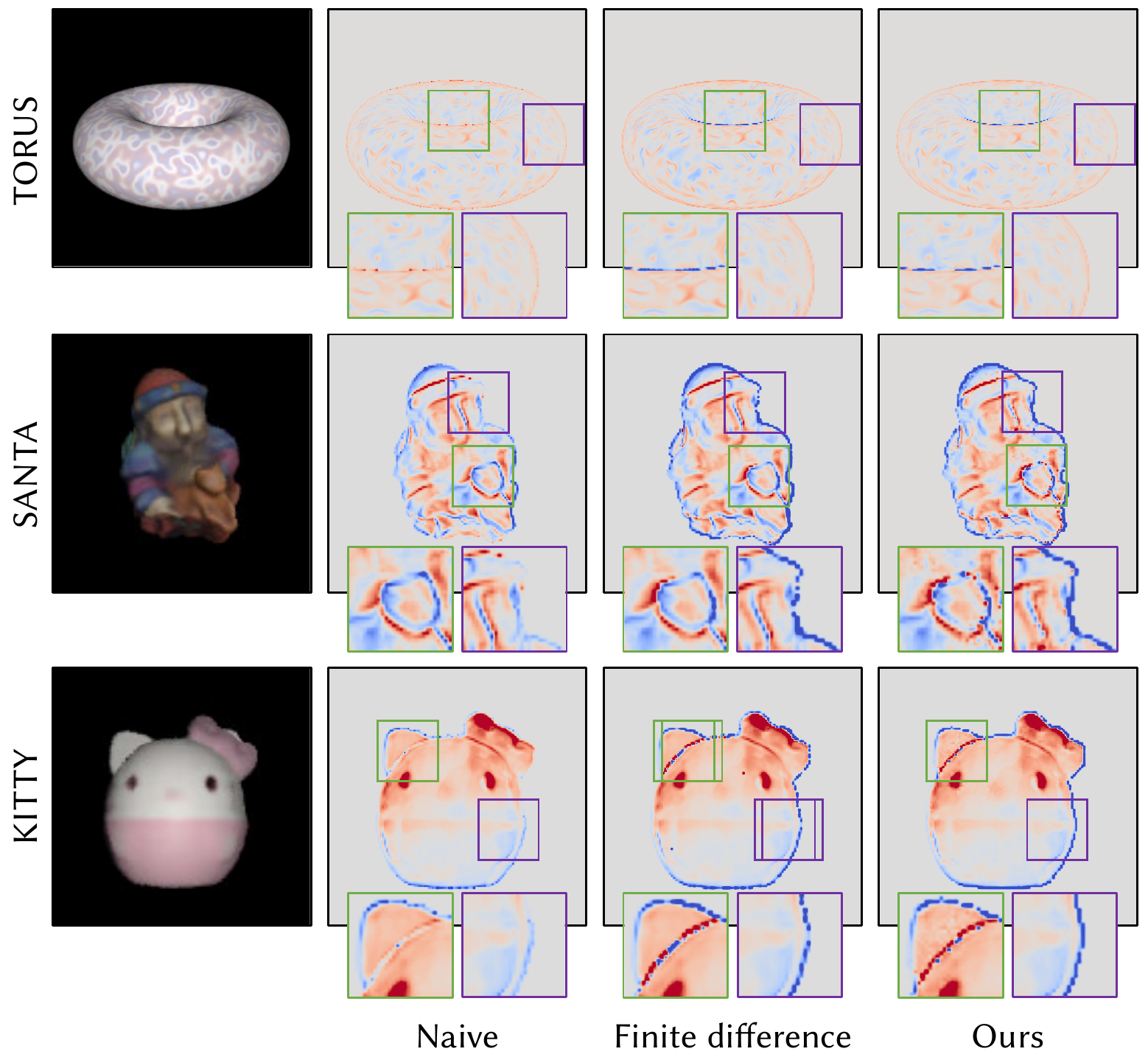}
\vspace{-.25in}
\caption{\textsc{Gradient quality}. We compare the image gradients computed na\"ively without reparameterization~\cite{Yariv:2020:IDR} and with our method against the ``ground truth'' gradient computed with finite differences for three scenes. Our method properly handles boundary discontinuities both due to object edges (in purple insets) and self-occlusions (in green insets).}
\label{fig:grad-quality}
\end{figure}

\subsection{Ground truth gradient comparions}
We first evaluate the correctness of our gradient by visualizing gradients on three different scenes (illustrated in Fig.~\ref{fig:grad-quality}). 
For \textsc{Torus}---a analytical torus model textured with a diffuse Perlin noise albedo---we visualize the gradients w.r.t the outer radius (distance from the center to the center of the ring). 
(\textsc{Santa} and \textsc{Kitty}) are 3D models that we represent as neural SDFs. 
We take the parameters of the neural SDF from an intermediate iteration during an inverse rendering optimization, and visualize the gradient w.r.t the bias parameter of the last layer output (i.e. the level set perturbation).
We also compute the gradient without reparameterization; this is similar to the gradient used in previous SDF-based inverse rendering methods~\cite{Yariv:2020:IDR}.
Note that the interior gradient is largely unaffected by reparameterization, with the gradient at the silhouettes being the largest benefit of our method, especially at self-occlusions. In the next subsection, we show that this boundary gradient is critical and without it, the inverse rendering diverges.

\begin{figure*}[th]
\centering
\setlength{\tabcolsep}{1pt}
{
\renewcommand{\arraystretch}{0.6}
\begin{tabular}{ccccccc}
\raisebox{40pt}{\rotatebox[origin=c]{90}{\textsc{santa}}} &
\includegraphics[width=0.16\linewidth]{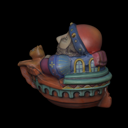} &
\includegraphics[width=0.16\linewidth]{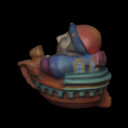} &
\includegraphics[width=0.16\linewidth]{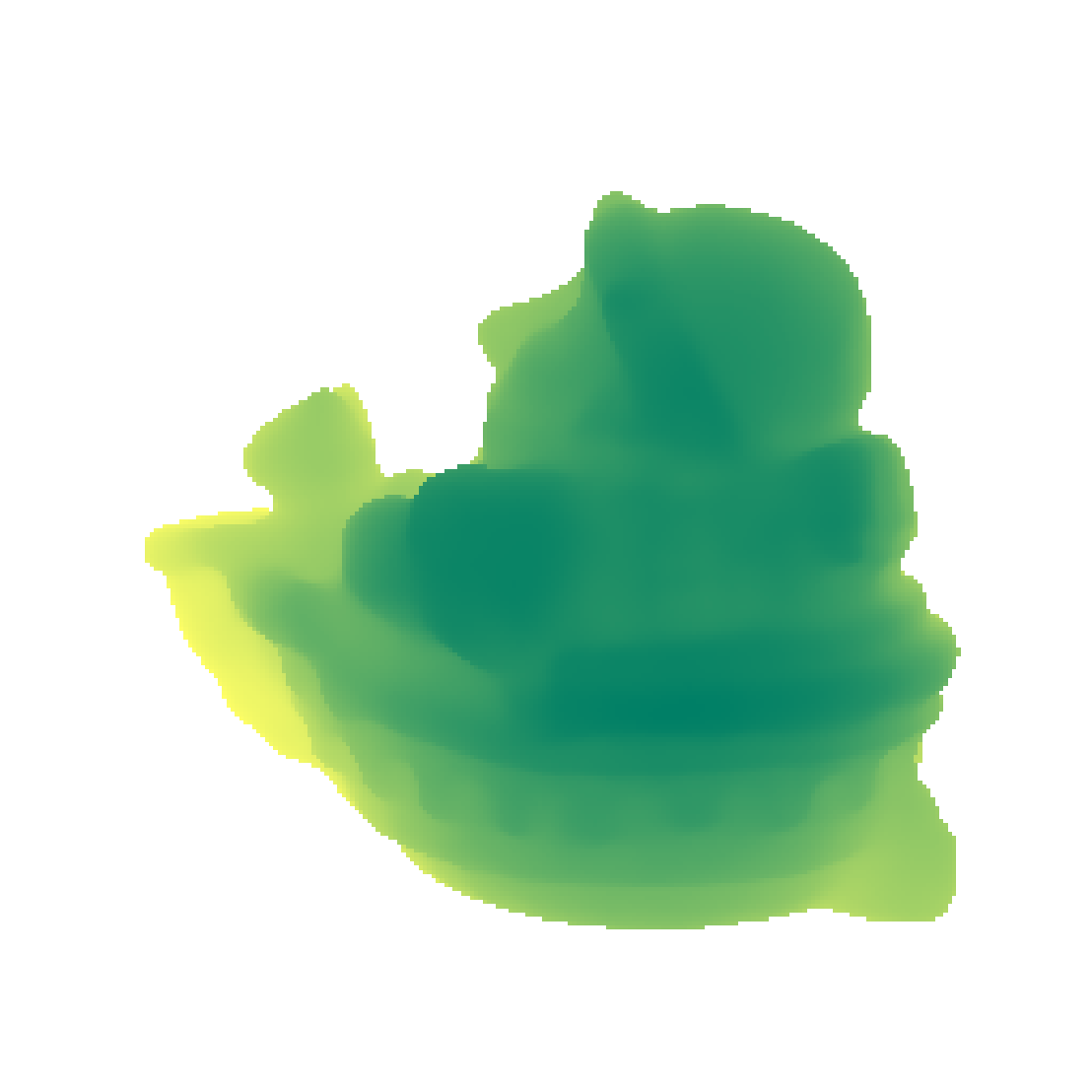} &
\includegraphics[width=0.16\linewidth]{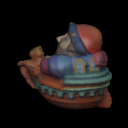} &
\includegraphics[width=0.16\linewidth]{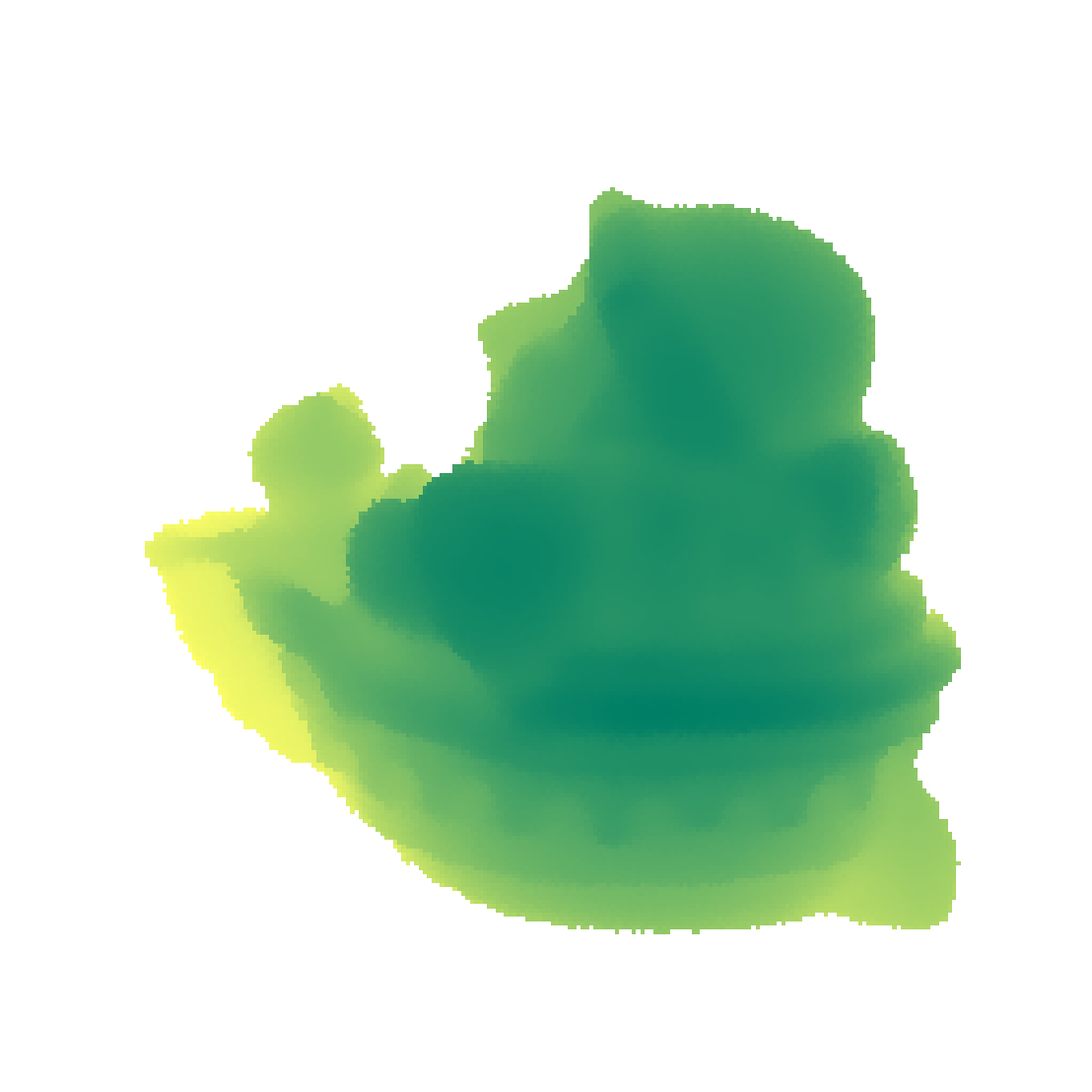} & 
\includegraphics[width=0.16\linewidth]{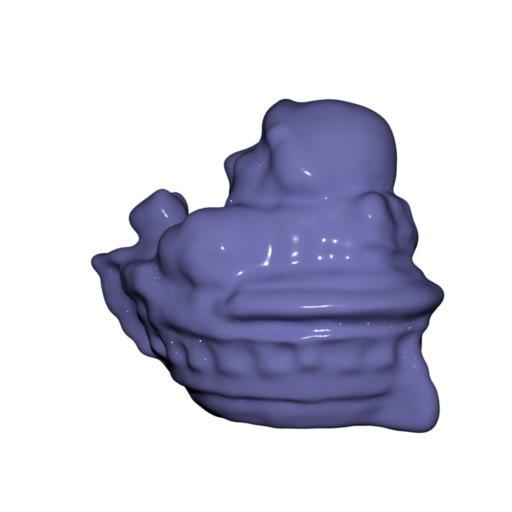} \\
\raisebox{40pt}{\rotatebox[origin=c]{90}{\textsc{kitty}}} &
\includegraphics[width=0.16\linewidth]{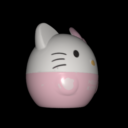} &
\includegraphics[width=0.16\linewidth]{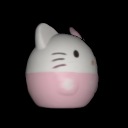} &
\includegraphics[width=0.16\linewidth]{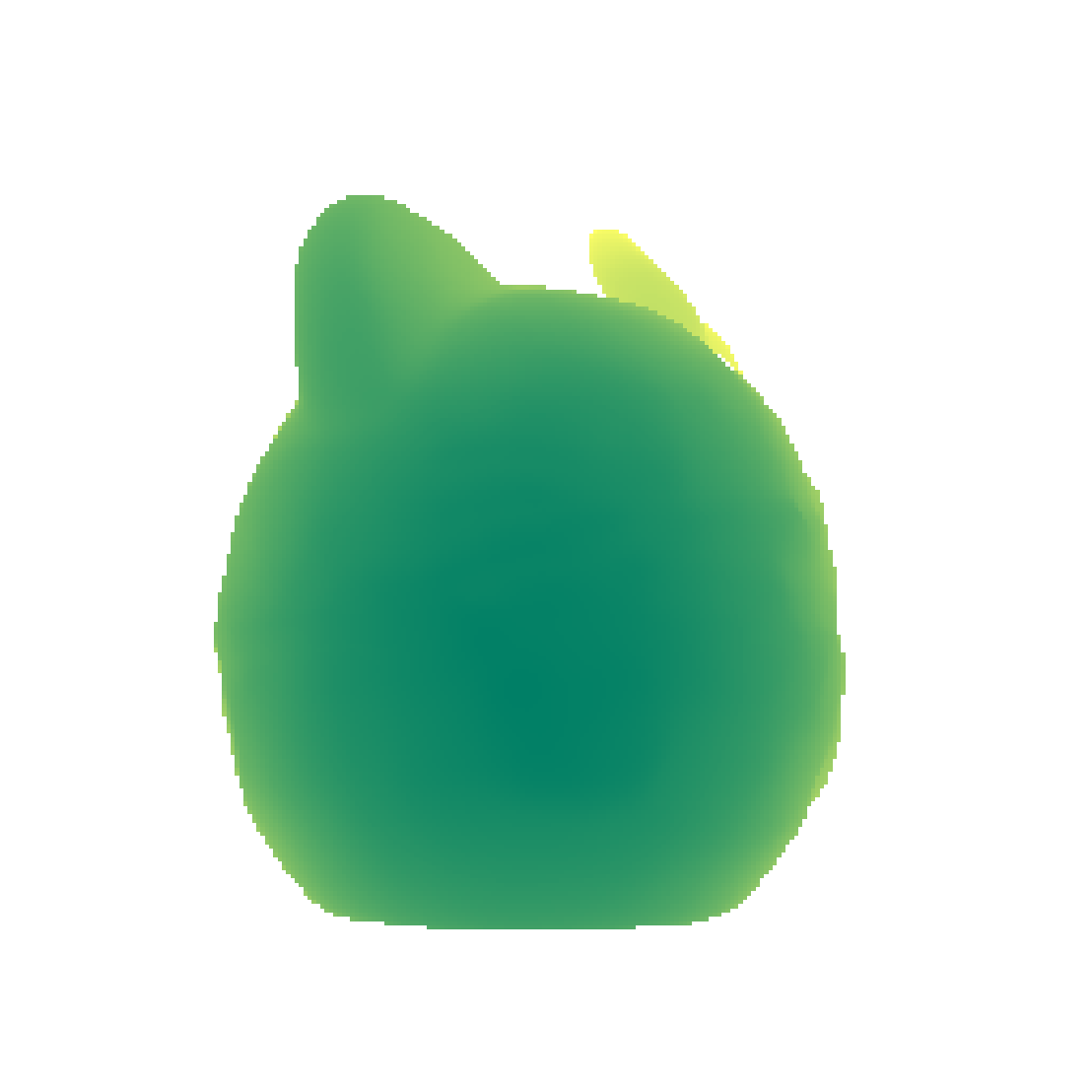} &
\includegraphics[width=0.16\linewidth]{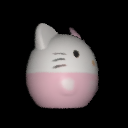} &
\includegraphics[width=0.16\linewidth]{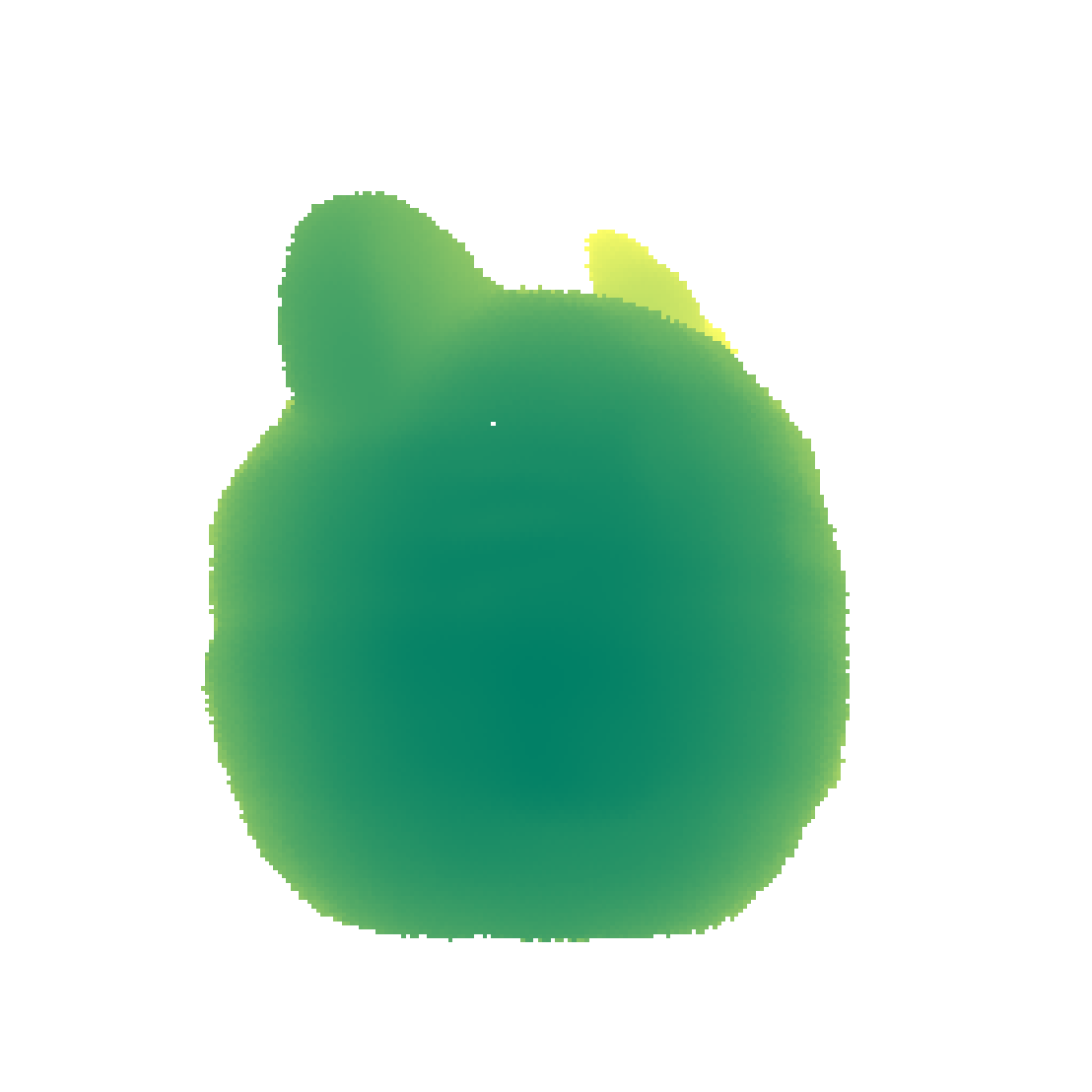} & 
\includegraphics[width=0.16\linewidth]{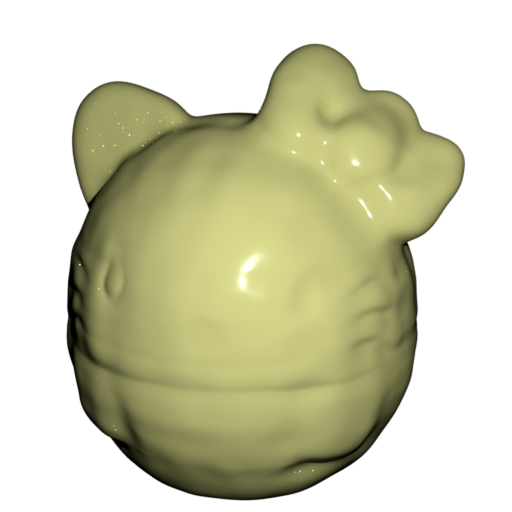} \\
\raisebox{40pt}{\rotatebox[origin=c]{90}{\textsc{duck}}} &
\includegraphics[width=0.16\linewidth]{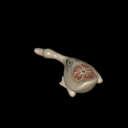} &
\includegraphics[width=0.16\linewidth]{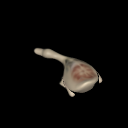} &
\includegraphics[width=0.16\linewidth]{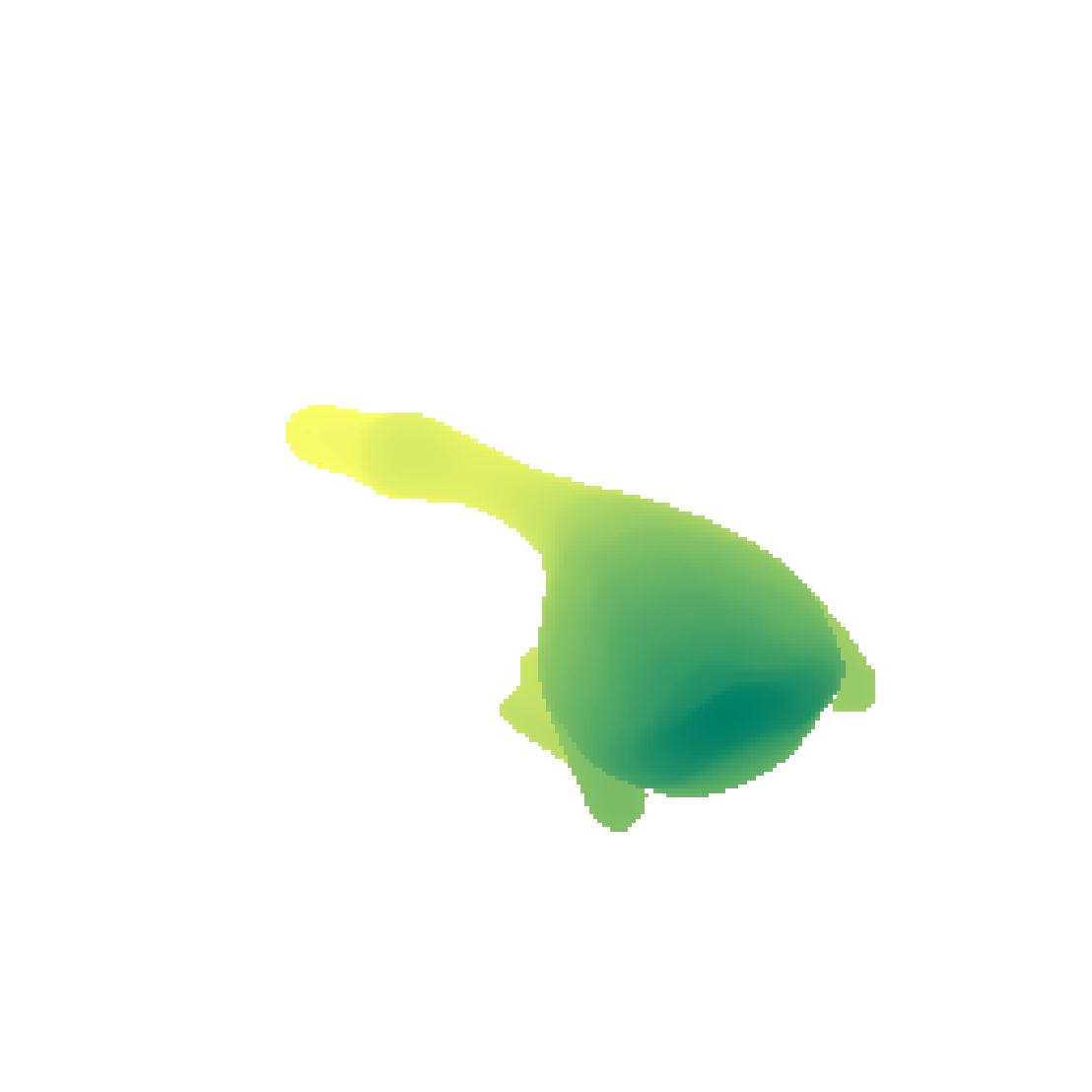} &
\includegraphics[width=0.16\linewidth]{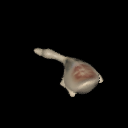} &
\includegraphics[width=0.16\linewidth]{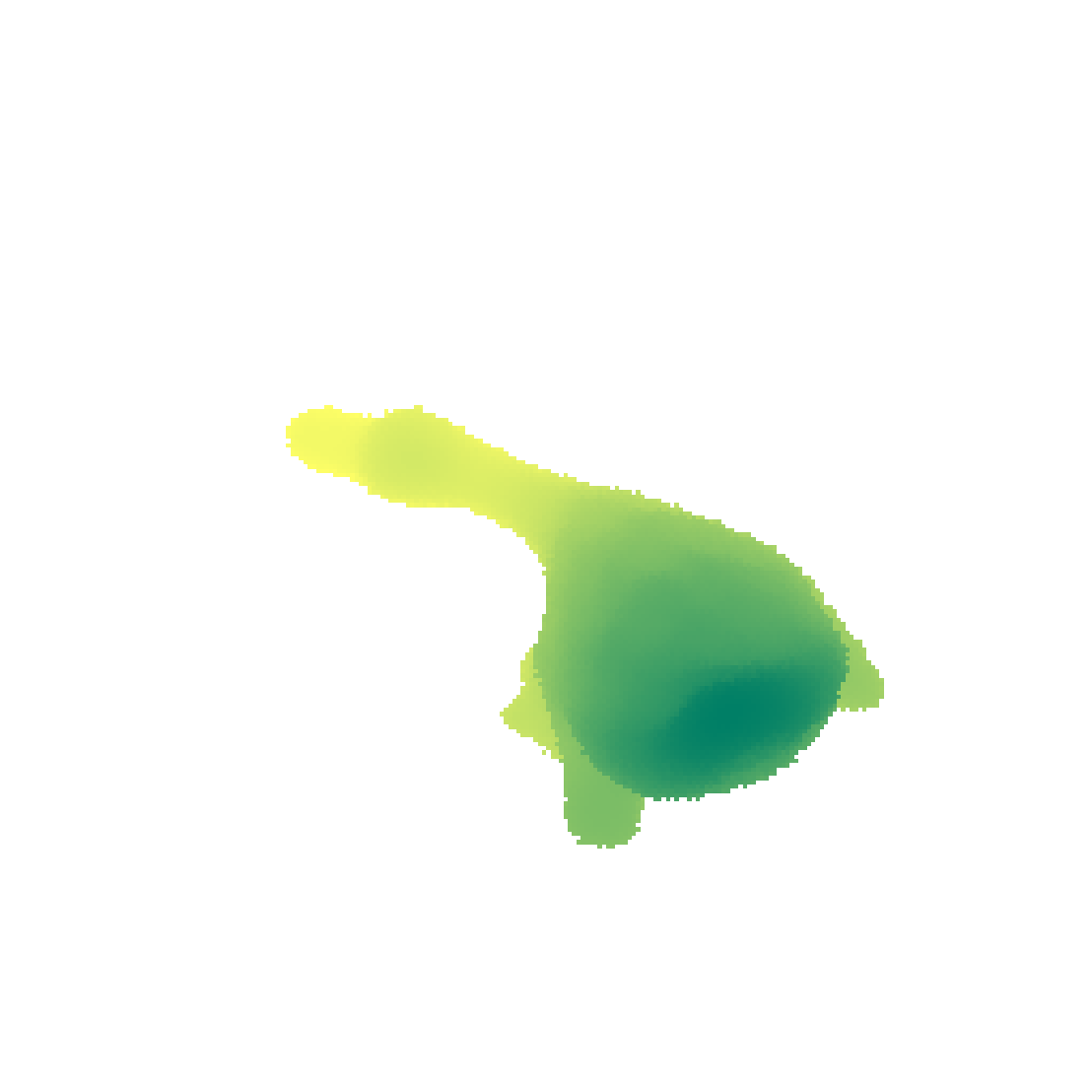} &
\includegraphics[width=0.16\linewidth]{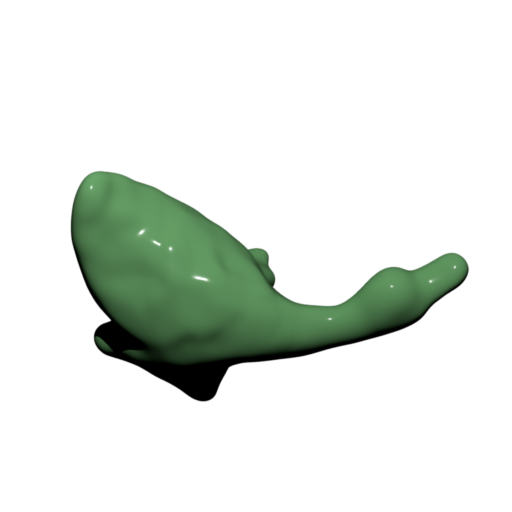} \\
\raisebox{40pt}{\rotatebox[origin=c]{90}{\textsc{pony}}} &
\includegraphics[width=0.16\linewidth]{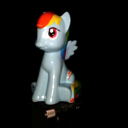} &
\includegraphics[width=0.16\linewidth]{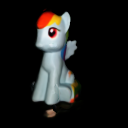} &
\includegraphics[width=0.16\linewidth]{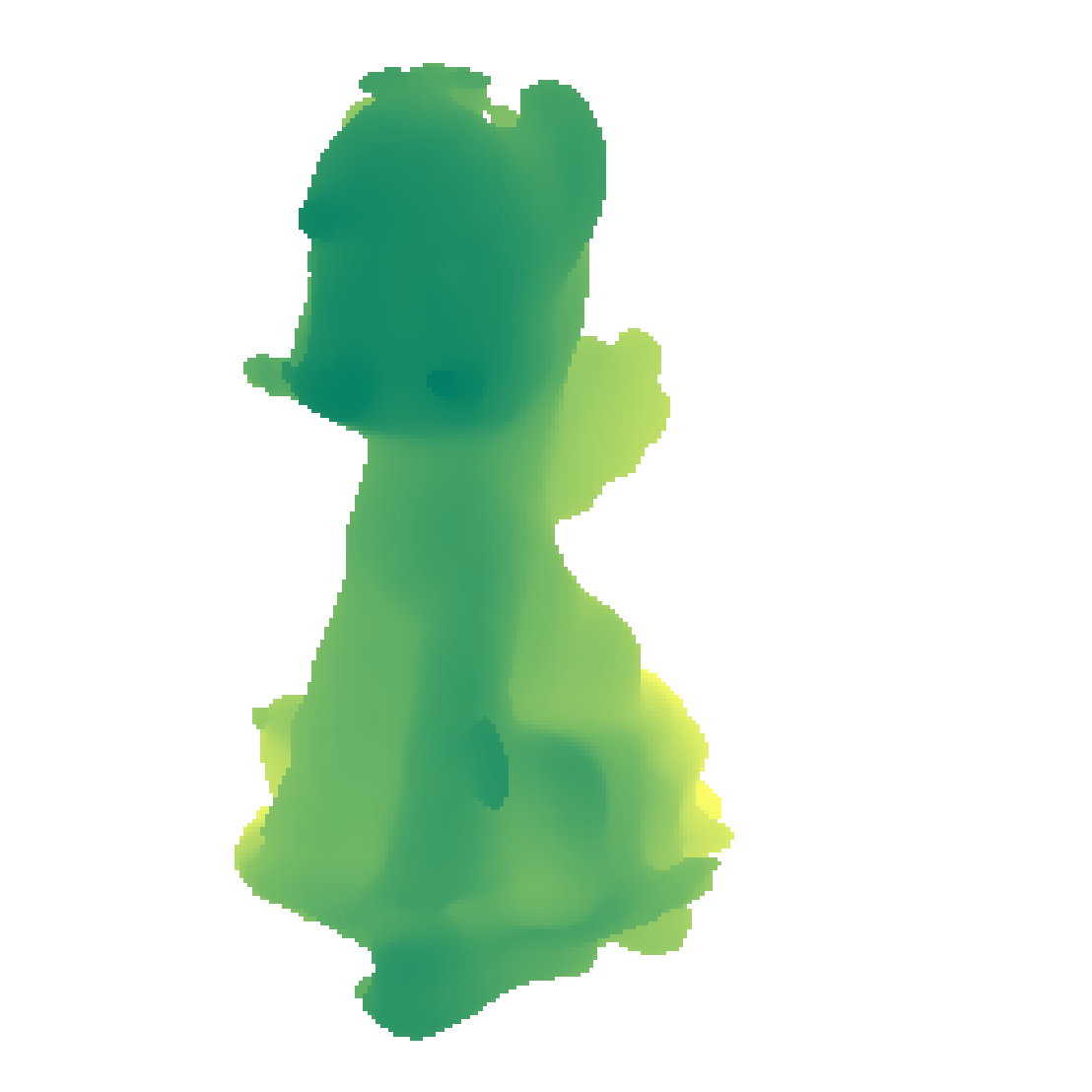} &
\includegraphics[width=0.16\linewidth]{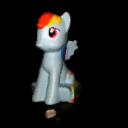} &
\includegraphics[width=0.16\linewidth]{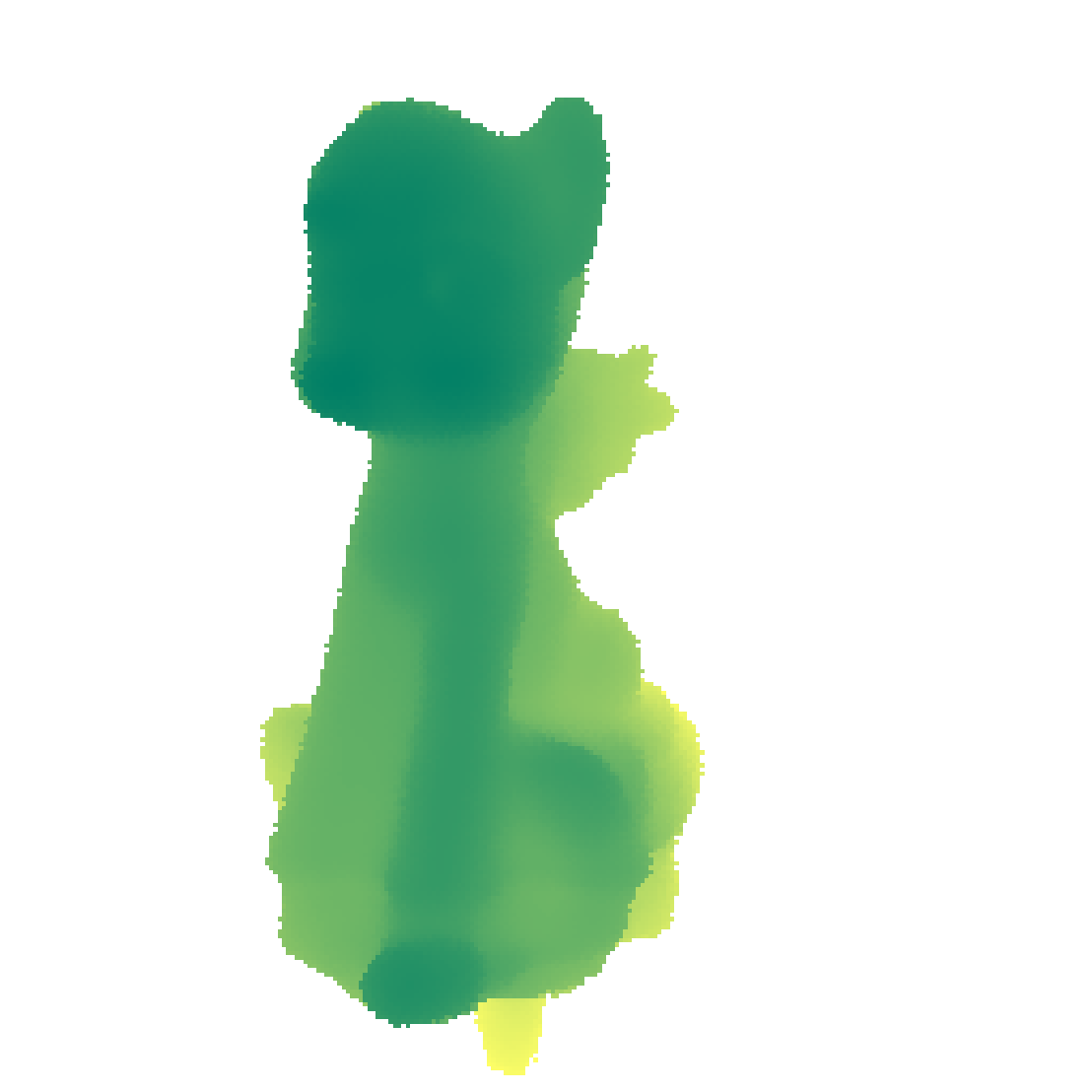} & 
\includegraphics[width=0.16\linewidth]{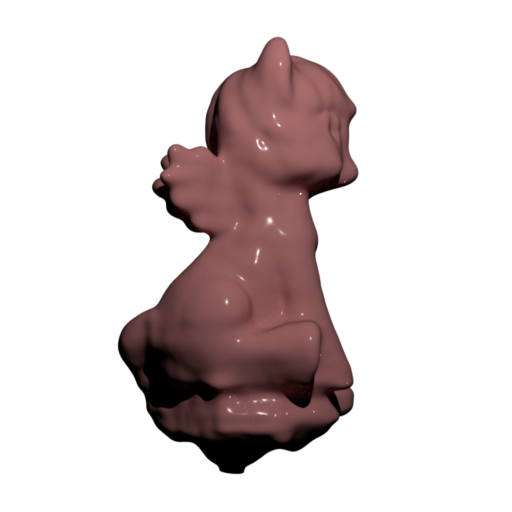} \\
\raisebox{40pt}{\rotatebox[origin=c]{90}{\textsc{dragon}}} &
\includegraphics[width=0.16\linewidth]{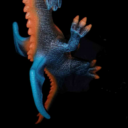} &
\includegraphics[width=0.16\linewidth]{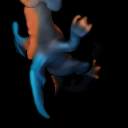} &
\includegraphics[width=0.16\linewidth]{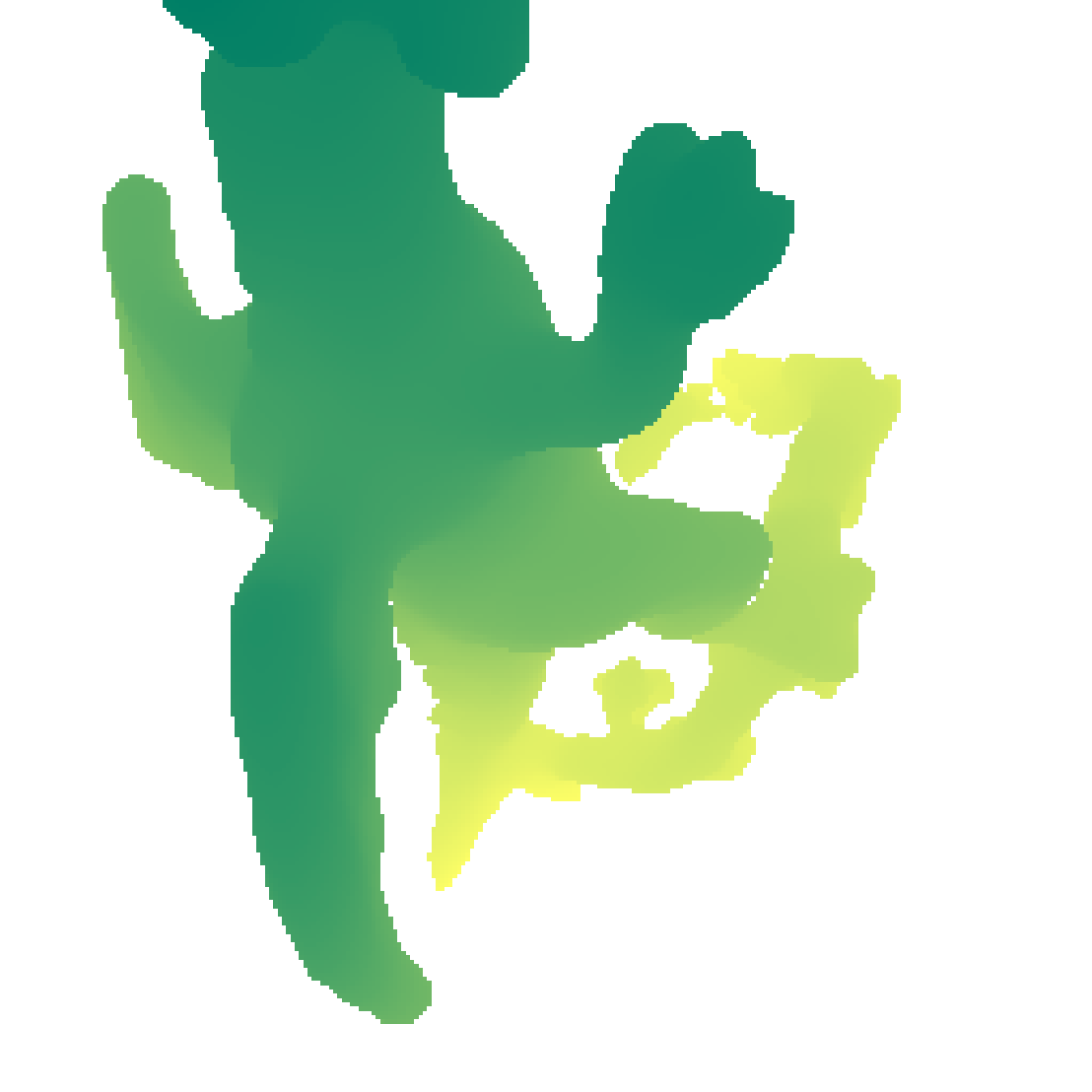} &
\includegraphics[width=0.16\linewidth]{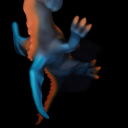} &
\includegraphics[width=0.16\linewidth]{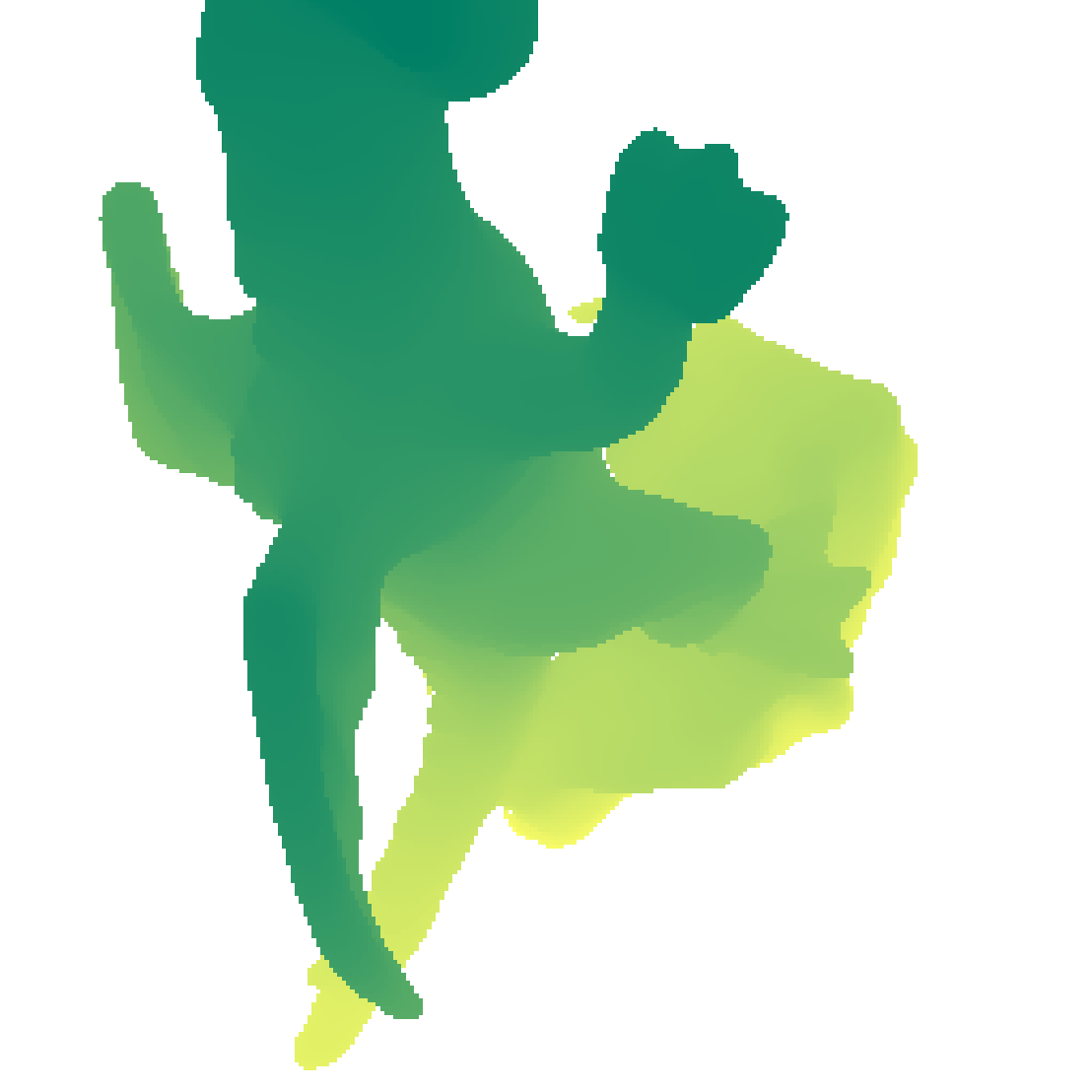} & 
\includegraphics[width=0.16\linewidth]{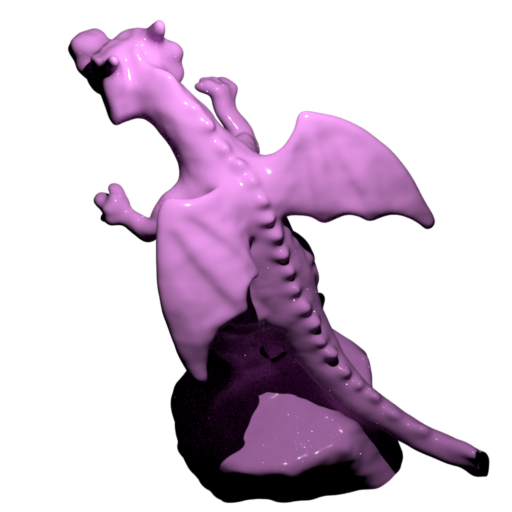}\\
& \textsc{Input image} & \textsc{IDR Radiance} & \textsc{IDR Depth} & \textsc{Ours Radiance} & \textsc{Ours Depth} & \textsc{Ours 3D geometry} \\
\end{tabular}
}
\vspace{-.1in}
\caption{\label{fig:neural-sdf-recon}\textsc{Neural SDF reconstruction.} We compare  with IDR~\cite{Yariv:2020:IDR} on three synthetic scenes (top three rows) and two real captured scene (bottom rows).
IDR requires 2D mask supervision, without which it completely diverges.
Thanks to our accurate gradient computation, our reconstructions are on par with IDR, without requiring any additional supervision beyond the input images and camera poses. In fact, on the real scenes, our reconstructions (without masks) outperform IDR \emph{with} masks (see the head of the \textsc{pony}, or the legs, tails and wings of the \textsc{dragon}) because of errors in automatic 2D segmentation.}
\end{figure*}
\subsection{Comparisons against IDR}
We compare our reconstructions against the SDF-based inverse rendering method of IDR~\cite{Yariv:2020:IDR}. IDR does not correctly account for the boundary term of gradient of the rendering integral and requires additional supervision, in the form of accurate 2D segmentation masks. We implement IDR in our pipeline to ensure that the only difference is our reparameterization. We use the same network architecture for both methods (See Sec. \ref{sec:inverse_rendering} for details), and report results after roughly 25,000 network updates. Note that our method uses more samples ($2$ in the interior + $4$ on each pixel boundary) since we use a Monte-Carlo approach to estimate the warp. IDR only requires one sample, fixed at the center of the pixel.

Figure~\ref{fig:neural-sdf-recon} shows that, on three synthetic scenes (\textsc{Santa}, \textsc{Kitty} and \textsc{Duck}), our method \emph{without} any 2D masks supervision obtains comparable depth and RGB reconstruction as IDR \emph{with} (perfect) mask supervision. We also show reconstructions of a captured real scene (\textsc{Pony} from ~\citet{Bi:2020:DRV}). Here, we provide IDR with 2D masks derived from a COLMAP reconstruction, which has errors. As a result, our reconstruction outperforms IDR on this scene.

We also tried to compare with IDR without mask supervision. In most cases, IDR without masks diverges completely because of the lack of gradients from the silhouette. This is similar to the observation made by \citet{Oechsle2021ICCV}.

\subsection{Ablation study: Subset size}
Our top-k weighting scheme reduces the memory footprint of our optimization, but this comes at a cost.
The smaller $k$, the sharper the weight landscape.
This can cause high variance that can impede the optimization of fine details.
We explore this through an ablation study 
on the \textsc{Santa} dataset, varying $k$ shown below. We use 36 views for this study, and report results after $20,000$ network updates.
%
Details are resolved for $k \geq 14$.

\vspace{.1in}
{
\centering
\setlength{\tabcolsep}{1pt}
\begin{tabular}{cccc}
\includegraphics[width=0.25\linewidth]{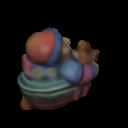} &
\includegraphics[width=0.25\linewidth]{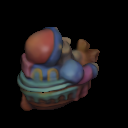} &
\includegraphics[width=0.25\linewidth]{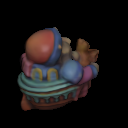} &
\includegraphics[width=0.25\linewidth]{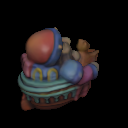}\\
$k=4$ & $k=8$ & $k=15$ & $k=22$
\end{tabular}
}

\section{Conclusion}
We have presented a novel method to correctly differentiate neural SDFs. 
Unlike prior work that relies on accurate masks or biased approximations of the boundary gradients, we reparameterize the pixel coordinates with discontinuity-aware transformation such that geometric derivatives exist everywhere and can be correctly computed via standard automatic differentiation. 
We have validated the correctness of our approach by comparing our gradients with finite difference ground truth, and demonstrated superior optimization convergence comparing state-of-the-art neural SDF baseline. 

While we have focused on primary visibility in this work, our formulation can be extended to global light transport. 
In particular, we expect to be able to model light rays and jointly optimize for geometry (represented as an SDF) as well as surface reflectance (instead of the radiance we are currently reconstructing) and illumination.
Modeling full global light transport (interreflections) with neural SDFs will require extensions or approximations to be computationally tractable.
Finally, inverse rendering under unknown, natural illumination is ill-posed and it would be interesting to explore geometry, material and illumination priors that can be combined with our differentiable rendering formulation.

\begin{acks}
This work was partially completed during an internship at Adobe Research and subsequently funded by the Toyota Research Institute and the National Science Foundation (NSF 2105806).
\end{acks}

\clearpage
\bibliographystyle{ACM-Reference-Format}
\bibliography{refs}

\appendix
\section{Pixel boundary sampling}
\label{sec:pixel_boundary_sampling}
\begin{figure}[!tb]
    \centering
    \includegraphics[width=\columnwidth]{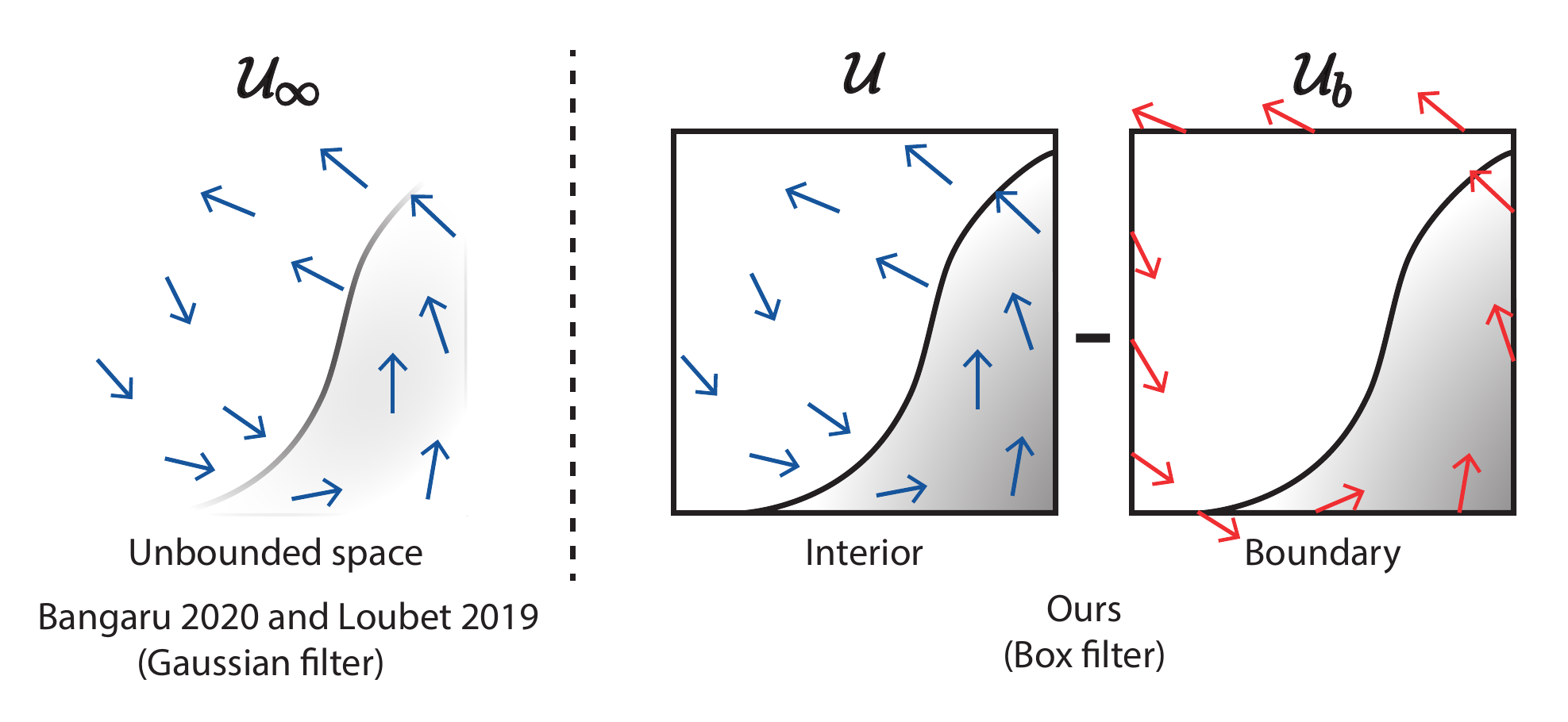}
    \vspace{-.25in}
    \caption{\label{fig:pixel-boundary} On the left, existing methods~\cite{Bangaru:2020:WAS, Loubet:2019:RDI} use an unbounded Gaussian filter to avoid the need to handle the boundary of the pixel filter $\mathcal{U}_b$, but this comes at the cost of increased variance in the interior due to the derivative of the Gaussian weights.
    On the right, our method uses a box filter and explicitly removes the  discrepancy in the warp field $\mathcal{V}$ through a boundary integral over $\mathcal{U}_b$.}
\end{figure}

In Eqn. \ref{eqn:divergence_theorem}, we use a box which implies the pixel domain $\mathcal{U}$ is \textit{bounded}. An implication of this is that we must also consider the boundaries of the pixel filter support (denoted by $\mathcal{U}_{b} \subset \mathtt{R}^2$) as discontinuities in $\mathcal{U}_{\text{sil}}$. Previously, to avoid this additional complexity, \citet{Bangaru:2020:WAS} used a Gaussian filter that has infinite support. 
We have found that this introduces extra variance due to the variation in the pixel filters in the divergence.

We instead keep the box filter as well as exclude the pixel boundary from the area integral $\mathcal{U}_{\text{sil}}$.
This means that Eqn. \ref{eqn:divergence_theorem} is no longer valid since the product $(L\mathcal{V})$ does not vanish smoothly at the pixel filter boundary $\mathcal{U}_{\text{b}}$.
We must instead rewrite the integral domain as an unbounded space $\mathcal{U}_{\infty}$.
We can further split the unbounded integral into two parts, one inside the pixel filter domain $\mathcal{U}$ and one outside (we omit parentheses here for brevity)
\begin{equation}
    I_{\text{sil}} = \int_{\mathcal{U}\setminus\mathcal{U}_{\text{sil}}} \nabla \cdot \left(L\mathcal{V}\right) + \int_{(\mathcal{U}_{\infty}\setminus\mathcal{U})\setminus\mathcal{U}_{\text{b}}} \nabla \cdot \left(L\mathcal{V}\right).
\end{equation}
We can then use the divergence theorem on the second area integral to turn it into a boundary integral over $\mathcal{U}_{\text{b}}$
\begin{equation}
    I_{\text{sil}} = \int_{\mathcal{U}\setminus\mathcal{U}_{\text{sil}}} \nabla \cdot \left(L\mathcal{V}\right) - \oint_{\mathcal{U}_{\text{b}}}L \left(\mathcal{V} \cdot \mathbf{n}_b \right),
\end{equation}
where $\mathbf{n}_b$ is the outward pointing normal of the pixel filter boundary, and the negative sign comes from the fact that we consider regions outside of the pixel filter instead of inside.
Unlike silhouette boundaries in $\mathcal{U}_{\text{sil}}$, $\mathcal{U}_b$ is easy to sample since it only contains axis-aligned line segments of equal length.
Fig. \ref{fig:pixel-boundary} illustrates the difference between using a smooth unbounded filter and using a box filter with pixel boundary sampling.

\end{document}


\maketitle

\newtheorem{theorem}{Theorem}[section]
\newtheorem{corollary}{Corollary}[theorem]
\newtheorem{lemma}[theorem]{Lemma}
\newtheorem{assumption}{Assumption}

\section{Correctness Sketch of $\mathcal{V}^{\text{q}}$}
\label{sec:v-q-correctness}
To show that $\mathcal{V}^{q}$ is valid we need to show that it is (i) Continuous and (ii) Boundary consistent.  
Here, we show that our weights are correct for an ideal $C_1$ continuous SDF and for an ideal sphere tracer $\mathcal{T}(u)$. Here, $\mathcal{T}(u)$ denotes the infinite series of points generated by the sphere tracer. Note that, in general, none of these points will actually satisfy $f(\mathbf{x}) = 0$ since an ideal sphere tracer never reaches the surface of an ideal SDF. Instead we will deal in \textit{limits}. That is,  $\lim_{n\to \infty}f(\mathbf{x}_n) = 0, \text{ } \mathbf{x}_n \in \mathcal{T}(\mathbf{u})$

\begin{assumption}
\textsc{(General Position Assumption)}. For a given $u$, there exist no two points $\mathbf{x}_i, \mathbf{x}_j \in \mathcal{T}(u), i \neq j$ such that both $f(\mathbf{x}_i) = f(\mathbf{x}_j)$ and $\partial_{\mathbf{x}} f(\mathbf{x}_i) = \partial_{\mathbf{x}} f(\mathbf{x}_j)$.
\label{lemma:general-position}
\end{assumption}

\begin{figure}
    \centering
    \includegraphics{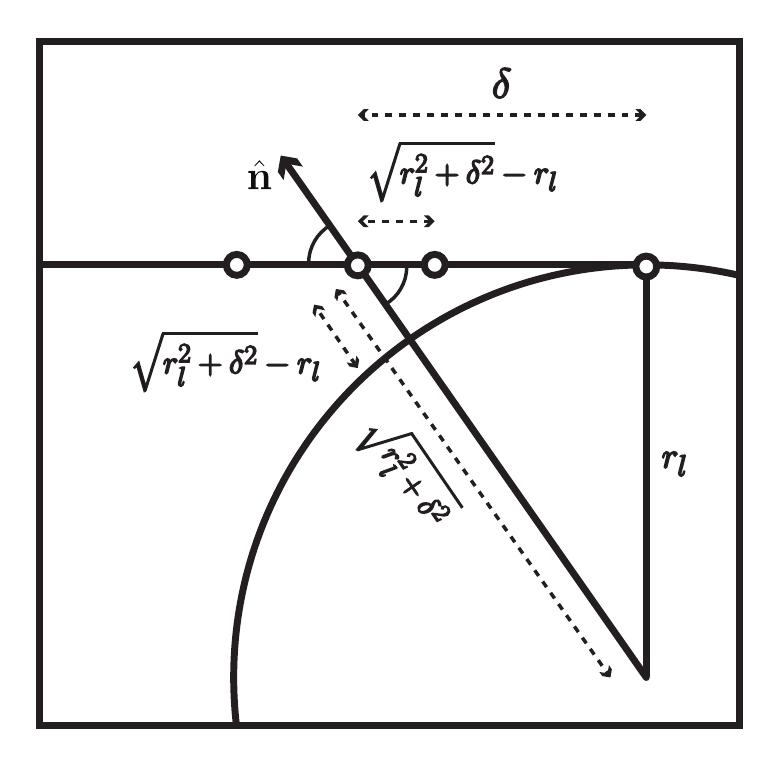}
    \caption{\textsc{Circle SDF Bound}}
    \label{fig:circle-sdf-bound}
\end{figure}
\begin{lemma}
\textsc{(Spherical Lower Bound)}.
There is an $\epsilon$-neighbourhood around every 3D silhouette point $\mathbf{x}_{\text{sil}}$, such that the SDF  $f(\mathbf{x})$ can be lower bounded by the SDF of a sphere with some fixed radius $r_l$ 
\label{lemma:sphere-lower-bound}
\end{lemma}
Since we can choose both $\epsilon$ and $r_l$, we can reduce the former and increase the latter until this lemma is satisfied.
The only way this scheme fails is if the curvature of the surface is 0 at $\mathbf{x}_{\text{sil}}$. That cannot be the case, because then the surface would be a plane parallel to the ray direction, which means all points along the ray contradict Assumption \ref{lemma:general-position} 

\begin{lemma}
\textsc{(Weight Lower Bound)}.
For a quadrature point along the silhouette ray that is distance $\delta$ away from the silhouette point, the weights can be lower bounded.
\begin{equation}
\lim_{\mathbf{u} \to \mathcal{U}_{\text{sil}}} w^{\text{q}}(\mathbf{x}(\mathbf{u}, t_{\text{sil}} - \delta)) \geq (\sqrt{r_l^2 + \delta^2} - r_l + \frac{\delta}{\sqrt{r_l^2 + \delta^2}})^{-\gamma} \cdot (\sqrt{r_l^2 + \delta^2} - r_l)
\end{equation}
\label{lemma:wt-q-lower-bound}
\end{lemma}
Fig. \ref{fig:circle-sdf-bound} illustrates the slice of the sphere SDF that contains the center of the sphere and the ray direction. For a point $\delta$ away from the silhouette point, the sphere SDF at $f(\mathbf{x})$ is $\sqrt{r_l^2 + \delta^2} - r_l$. 
Using the property of similar triangles, the dot product of the normal with the direction is $\delta / \sqrt{r_l^2 + \delta^2}$. 

\begin{lemma}
\textsc{(Unboundedness of the Lower Bound)}.
For $\gamma > 2$, the lower bound in Lemma \ref{lemma:wt-q-lower-bound} is unbounded in the limit $\delta \to 0$
\begin{equation}
\lim_{\delta \to 0}(\sqrt{r_l^2 + \delta^2} - r_l + \lambda_d \cdot  \frac{\delta}{\sqrt{r_l^2 + \delta^2}})^{-\gamma} \cdot (\sqrt{r_l^2 + \delta^2} - r_l) = \infty
\end{equation}
\label{lemma:unbounded-lower-bound}
\end{lemma}

To see this, notice that the limit above can be written as
\begin{equation}
    \lim_{\delta \rightarrow 0} \frac{\sqrt{r_l^2+\delta^2}-r_l}{\left(\sqrt{r_l^2 + \delta^2} - r_l + \lambda_d \cdot \frac{\delta}{\sqrt{r_l^2 + \delta^2}}\right)^{\gamma}}.
\end{equation}
Taking the Taylor expansion at $\delta=0$, for the numerator we have:
\begin{equation}
    \sqrt{r_l^2+\delta^2}-r_l = \frac{\delta^2}{2r_l} + O(\delta^4).
\end{equation}
For the denominator we have:
\begin{equation}
    \left(\sqrt{r_l^2 + \delta^2} - r_l + \lambda_d \cdot \frac{\delta}{\sqrt{r_l^2 + \delta^2}}
    \right)^{\gamma} = \frac{\delta^{\gamma}}{r_l} +  O(\delta^{\gamma + 1})
\end{equation}
Substituting, we have that the ratio is asymptotically equivalent to:
\begin{equation}
    \frac{\frac{\delta^2}{2r_l} + O(\delta^4)}{\frac{\delta^{\gamma}}{r_l} + O(\delta^{\gamma + 1})} =
    \frac{\frac{\delta^{2-\gamma}}{2r_l} + O(\delta^{4-\gamma})}{\frac{1}{r_l} + O(\delta)},
\end{equation}
which diverges as long as $\gamma > 2$ .

\begin{lemma}
\textsc{(Kronecker Delta Behaviour)}.
For a ray exactly at the silhouette, the limiting point of the sphere tracer is assigned all the weight, given $\gamma > 2$ and $\lambda_d > 0$
\begin{equation}
\lim_{\mathbf{u} \to \mathcal{U}_{\text{sil}}}
\lim_{\mathbf{n} \to \infty} \frac{w^{\text{(q)}}\left(\mathbf{x}_n(\mathbf{u};t)\right)}{\sum_{\mathbf{x}_i \in \mathcal{T}(u)} w^{\text{(q)}}\left(\mathbf{x}(\mathbf{u};t^{\prime})\right) \cdot \mathrm{d}t^{\prime}} = 1    
\end{equation}
\label{lemma:wts-kronecker-delta}
\end{lemma}
We can show this through contradiction. 
Since the number of sphere tracer points are countably infinite, let's consider some point $\mathbf{x}_{i} \in \mathcal{T}(u)$ that is not the limiting point. 
From \ref{lemma:wts-kronecker-delta} and \ref{lemma:unbounded-lower-bound},
since $\mathcal{T}(u)$ is an infinite series, we can necessarily find a point $\mathbf{x}_j, j > i$ such that $\frac{w^{\text{(q)}}(\mathbf{x}_i)}{w^{\text{(q)}}(\mathbf{x}_j)} < p$ for \textit{any} $p > 0$
Therefore, in the limit of $n \to \infty$, the normalized weight of $\mathbf{x}_i$ is 0.

This is true for every point $x_i \in \mathcal{T}(u)$ that is not the limiting point itself. Thus, our weights become a discrete version of delta (i.e. the \textit{Kronecker delta}) on the limiting point.

Since the limiting point of $T(u_{\text{sil}})$ is $\mathbf{x}_{\text{sil}}$, it follows from the form of our quadrature weights that,
$ \lim_{\mathbf{u} \to \mathcal{U}_{\text{sil}}} \mathcal{V}^{\text{q}}(\mathbf{u}) = G(\mathbf{x}(u, t_{\text{sil}});\theta)^T\partial_{\mathbf{x}}u$.
That is, $\mathcal{V}^q$ is boundary consistent.

\section{Correctness Sketch of top-k weights $\bar{w}_{k}$}

Since $T_k(u)$ always contains the $k$ points with the largest weights, boundary consistency follows from the correctness of $\mathcal{V}^{\text{(q)}}$. However, \textit{continuity} is non-trivial since the discrete set of points in $\mathcal{T}_k(u)$ can change as $u$ changes. We also only need to consider continuity at non-silhouette points since the resulting warp field is never evaluated exactly at silhouette points. 

\begin{lemma}
\textsc{Top-K Weight Continuity} The weights of the set $\mathcal{T}_k(u)$ are continuous for all $u \not\in \mathcal{U}_{\text{sil}}$
\end{lemma}

We analyze the weights of the points in the set $\mathcal{T}_k(u)$ under two separate cases
\begin{enumerate}
    \item Case 1: The indices of points in $\mathcal{T}_k(u)$ change in the infinitesimal neighbourhood around $u$. 
    
    First, note that because of Assumption \ref{lemma:general-position}, no two points in $\mathcal{T}(u)$ (and $\mathcal{T}_k(u)$ by extension) can have the same weight. Thus, in an infinitesimally-small neighbourhood, we can assume that there is only one $\mathbf{x}_i \in \mathcal{T}_k(u)$ that is replaced with a new point $\mathbf{x}_j \in \mathcal{T}(u)$, $\mathbf{x}_j \not\in \mathcal{T}_k(u)$, as we perturb $u$.
    In this neighbourhood, we can assert that:
    \begin{equation*}
        w^{\text{q}}(\mathbf{x}_i) = w^{\text{q}}(\mathbf{x}_j)   
    \end{equation*}
    We can also note that 
    \begin{equation*}
     w^{\text{q}}(\mathbf{x}_i) = \min_{\mathbf{x_m} \in \mathcal{T}_k(u)} w^{\text{q}}(\mathbf{x}_m)   
    \end{equation*}
    because by definition of the top-k subset, only the smallest weight is swapped out of the set. 
    However, remember that, from our definition of top-k weights, because we shift every weight by the smallest weight,
    the smallest weight in the subset is zero, i.e.:
    \begin{equation*}
        w^{\text{k}}(\mathbf{x}_i) = w^{\text{k}}(\mathbf{x}_j) = 0
    \end{equation*}
    
    Therefore, because both the swapped points $\mathbf{x}_i$ and $\mathbf{x}_j$ have a weight of 0, the weights of $\mathcal{T}_k(u)$ are continuous in the neighbourhood of $u$.
    
    \item Case 2: The order of points in $\mathcal{T}_k(u)$ remain constant in the infinitesimal neighbourhood around $u$. 
    
    Since the points are at the same position in the original series $\mathcal{T}(u)$ and their weights are continuous, it follows that the weights of points in the subseries $\mathcal{T}_k(u)$ are also continuous.
\end{enumerate}